\documentclass[aps,prx,superscriptaddress,longbibliography,twocolumn]{revtex4}
\usepackage{enumitem}
\usepackage{graphicx,amssymb,amsmath,amssymb,natbib,bm}
\usepackage{epstopdf}
\usepackage[caption=false]{subfig}
\usepackage{float}
\usepackage{mathscinet}
\usepackage{color}
\newcommand{\n}{{\mathbf{n}}}

\newcommand{\x}{{\mathbf{x}}}
\newcommand{\bphi}{\pmb{\phi}}

\begin{document}
\title{Binary-state dynamics on complex networks: Stochastic pair approximation and beyond}
\author{A. F. Peralta}
\email{afperalta@ifisc.uib-csic.es}
\affiliation{IFISC (Instituto de F{\'\i}sica Interdisciplinar y Sistemas Complejos), Universitat de les Illes Balears-CSIC, 07122-Palma de Mallorca, Spain}
\author{R. Toral}
\affiliation{IFISC (Instituto de F{\'\i}sica Interdisciplinar y Sistemas Complejos), Universitat de les Illes Balears-CSIC, 07122-Palma de Mallorca, Spain}
\date{\today}
\begin{abstract}
Theoretical approaches to binary-state models on complex networks are generally restricted to infinite size systems, where a set of non-linear deterministic equations is assumed to characterize its dynamics and stationary properties. We develop in this work the stochastic formalism of the different compartmental approaches, these are: approximate master equation (AME), pair approximation (PA) and heterogeneous mean field (HMF), in descending order of accuracy. Using different system-size expansions of a general master equation, we are able to obtain approximate solutions of the fluctuations and finite-size corrections of the global state. On the one hand, far from criticality, the deviations from the deterministic solution are well captured by a Gaussian distribution whose properties we derive, including its correlation matrix and corrections to the average values. On the other hand, close to a critical point there are non-Gaussian statistical features that can be described by the finite-size scaling functions of the models. We show how to obtain the scaling functions departing only from the theory of the different approximations. We apply the techniques for a wide variety of binary-state models in different contexts, such as epidemic, opinion and ferromagnetic models. 
\end{abstract}
\keywords{Binary-state models, Pair approximation, Approximate master equation}
\maketitle

\section{Introduction}

Binary-state models on complex network are a very general theoretical framework to study the effect of interactions in the dynamics of a population of individuals. They are composed by a set of nodes that are connected between them through a particular random network, where each node holds a binary (``spin''-like, two values) variable that evolves in time by some given transition rates. Typical problems that can be mapped in this scheme include models of epidemic spreading \cite{Anderson:1991,Keeling:2007,Brauer:2011,Satorras:2015}, language competition \cite{Abrams:2003,Mira:2005,Castello:2009,Vazquez:2010,Patriarca:2012}, social interaction \cite{Galam:2002,Barrat:2008,Castellano:2009,Newman:2010,Castellano:2012,Nyczka:2013}, financial markets \cite{Alfarano:2005,Alfarano:2008,Alfarano:2009,Carro:2015,Vilela:2019,Kononovicius:2019}, among many others.

Recently, there has been a lot of effort in the development of highly accurate mathematical descriptions of the dynamics of these models. Typically, we can distinguish between two types of approaches depending on the variables that one chooses to describe the system: (i) individual based-approaches \cite{Hethcote:1984,Wang:2003,Chakrabarti:2008,Lafuerza:2013,Carro:2016}, where the ``spin" or state of each node of the network is considered as an independent variable, (ii) compartmental approaches \cite{Satorras:2001,Barrat:2008,Sood:2008,Vespignani:2012,Gleeson:2011,Gleeson:2013,Peralta_pair:2018}, where nodes sharing the same topological property such as, for example, the number of neighbors in the network, are aggregated in a single variable, being this an integer (occupation) number. Depending on the level of description, i.e. the number of variables and its nature, one distinguishes between different compartmental approaches: approximate master equation (AME) \cite{Marceau:2010,Lindquist:2011,Gleeson:2011,Gleeson:2013}, pair approximation (PA) \cite{Dickman:1986,Oliveira:1993,Vazquez:2008,Pugliese:2009,Schweitzer:2009,Mata:2014} and heterogeneous mean field (HMF) \cite{Satorras:2001,Sood:2008,Vespignani:2012}. Only the individual node and AME approaches can be considered as a complete description of the models, while the PA and HMF introduce constraints between variables which may or may not be fulfilled, thus they are generally a worse approximation. Except in a few cases where fluctuations are taken into account at some extent \cite{Ferreira:2011a,Ferreira:2011b} and in its completeness \cite{Lafuerza:2013,Carro:2016}, the approaches are usually followed by a deterministic type of description \cite{Satorras:2001,Boguna:2002,Wang:2003,Vespignani:2012}, where the stochastic nature of the models defined by the individual transitions rates is neglected. The deterministic approach enables one to obtain some important quantities of the models such as the critical point (e.g. the epidemic threshold), or the time evolution of the global state of the system (e.g. the density of infected population). The accuracy and suitability of the different approaches has been widely discussed in the literature. For example in the determination of the epidemic threshold, it has been shown that the two approaches, individual and compartmental, may give contradictory results \cite{Castellano:2010,Boguna:2013} and a general recipe for choosing one or another was given in \cite{Ferreira:2016}.

Although the deterministic approach gives us relevant information in all situations, it is accurate only in the strict infinite system size limit. Depending on the model, the variables chosen, the values of the parameters and the network, the difference between the deterministic approach and the numerical results may be very important in finite networks \cite{Castellano:2006}. Finite-size effects become relevant even for extremely large system sizes, specially if the system is close to a critical point, or the network has high degree heterogeneity. Besides, there are some types of models where the deterministic approach does not provide the relevant information sought. For example, the noisy voter (Kirman) model \cite{Fichthorn:1989,Considine:1989,Kirman:1993,Granovsky:1995} is an opinion model that considers neighbor imitation and random switching of opinion as basic ingredients. Different versions of the model have been applied in many different contexts, the most important in our perspective being the study price fluctuation in financial markets \cite{Alfarano:2005,Alfarano:2008} and vote share distributions in electoral data \cite{Gracia:2014,Kononovicius:2018,Kononovicius:2019b}. In this context, the global opinion does not take a fixed deterministic value but shows heavy fluctuations around the mean instead. The statistics of these fluctuations are the most important feature of study, as it shows deviations from the Gaussian behavior (the global variable distributes as a Beta distribution \cite{Kirman:1993,Alfarano:2005}) with very similar properties to financial series and vote share fluctuations. The model has a finite-size critical point that vanishes in the thermodynamic limit and thus a stochastic approach is mandatory in order to achieve the correct characterization \cite{Carro:2015,Peralta_pair:2018}. Additionally, the noisy voter model is of major importance because of its simplicity and the possibility of obtaining analytical results, which are helpful to fully understand its properties. Recent generalizations of the model include: the effect of non-linear copying mechanisms \cite{Jedrzejewski:2017,Peralta:2018,Vieira:2018}, non-Markovian memory effects \cite{Artime:2019,Oriol:2018a,Peralta:2020}, zealots \cite{Khalil:2018} and contrarians \cite{Khalil:2019}, more than two states \cite{Kononovicius:2018,Vazquez:2019,Herrerias:2019}, the role of different noise and copying mechanisms in the nature of the transition (continuous or discontinuous) \cite{Nowak:2019,Abramiuk:2019}, etc.

The main aim of this work is to give a general theoretical approach to binary-state models on complex networks that takes into account stochastic effects, going beyond simple incomplete deterministic approaches. With this intention we will first find the general master equation of the individual and compartmental approaches. The master equation corresponds to a full characterization of any Markovian process and one can derive easily the deterministic equations from it \cite{VanKampen:2007,Toral:2014}. Although the master equation of the individual and AME approaches give a very accurate result, they are hardly impossible to solve in most situations even computationally \cite{Economou:2015}. In order to overcome this issue and obtain at least an appropriate approximate solution, we will apply different expansion techniques of the master equation. The first one is a van Kampen-like system-size expansion \cite{VanKampen:2007,Grima:2012,Asslani:2012,Cianci:2017,Peralta_moments:2018}, where the variables are split between its deterministic value plus finite-size corrections. The van Kampen approach can be understood as an expansion in inverse powers of system size, and it is known to be accurate far from criticality with increasing accuracy when the system size increases \cite{Peralta_moments:2018}. As a natural extension of the van Kampen expansion we will also apply the Kramers-Moyal expansion \cite{VanKampen:2007} and derive the corresponding (continuous) Fokker-Planck equation from the original master equation. Close to a critical point, the finite-size effects of the models are well captured by their finite scaling functions \cite{Deutsch:1992,Castellano:2006,Ferreira:2011b,Mata:2014,Peralta:2018} which can not, a priori, be derived from the van Kampen expansion. The correct way of obtaining the theoretical scaling functions is to apply a similar system-size expansion of the master equation but with an anomalous scaling with system size \cite{Nakanishi:1990,Nakanishi:2000}. We will check the accuracy and suitability of the expansions as a function of the parameters for different models and networks.

The paper is organized as follows: in Section \ref{sec_general} we introduce the general definitions and notation of binary state models and the main characteristics of the networks. In Section \ref{sec_master} we construct the master equation for individual and compartmental approaches. In Section \ref{sec_master} we apply the van Kampen expansion to the general master equation and we show its connection to the Kramers-Moyal expansion. We re-derive the deterministic nonlinear equations \cite{Gleeson:2013}, together with a set of linear equations for the correlations and average values of the stochastic corrections. In Section \ref{sec_vKresults} we compare the results of the numerical simulations in the stationary state of small systems with the theoretical results of the van Kampen approach for different models and networks. In Section \ref{sec_critical} we apply the expansion of the master equation close to a critical point and compare it to the scaling functions obtained numerically in Section \ref{sec_critical_results}. In Section \ref{sec_time} we extend the numerical results to time dependent quantities, together with its comparison to the van Kampen results. We end with a summary and conclusions in Section \ref{sec_conclusion}. In the appendices we explain the intermediate steps in the derivation of the equations: in Appendix \ref{app:matrices} we write out the expressions of the matrices involved in the equations of the van Kampen and Kramers-Moyal expansions, while Appendix \ref{app:manifold} contains the details of the expansion around the critical point.

\section{general aspects, models and networks}\label{sec_general}

A binary-state model is composed of a population of size $N$, where each member of the population can be in two states $1$, ``adopter'', or $0$, ``non-adopter''. Depending on the model and the context, the states may represent different properties of the individuals, for example magnetic spin, opinion on a topic, spoken language, infection state, etc. This is naturally described by a set of time-dependent binary variables $\n(t) \equiv \lbrace n_{i}(t)=0,1 \rbrace_{i=1, \dots, N}$. Each individual $i=1, \dots, N$ is regarded as a node of a, single-connected, undirected network, which can be mapped into the usual (symmetric) adjacency matrix $\mathbf{A} = \lbrace A_{ij} \rbrace$, with coefficients $A_{ij}=1$ if nodes $i$ and $j$ are connected and $A_{ij}=0$ otherwise, where self-loops are avoided $A_{ii}=0$. The degree of node $i$ is defined as the total number of nodes connected to it $k_{i}=\sum_{j=1}^{N} A_{ij}$ (number of neighbors), taking values in between $k \in [k_{\text{min}}, k_{\text{max}}]$. The degree $k$ can be heterogeneous within the population and one defines the number $N_{k}$ of nodes with degree $k$, and the associated fraction $P_{k}=N_{k}/N$ called degree distribution. It is also useful to define the moments of degree \emph{m} as $\mu_{m}=\sum_{k \in [k_{\text{min}}, k_{\text{max}}]} P_{k} k^{m}$, with short notation $\mu_{1}=\mu$, which corresponds to the average degree. Networks are assumed to be generated by the configuration model \cite{Catanzaro:2005}, with fixed degree distribution $P_{k}$, which produces uncorrelated networks if $k_{\text{max}} \leq \sqrt{\mu N}$ (no degree-degree correlations and no triangles).

The dynamical model under study is defined by the individual rates $r_{i}^{\pm}$, which determine the time evolution of the state variables $n_{i}(t)$. They are defined as the probability per unit time that the transition $n_{i}=0 \rightarrow 1$ occurs, with rate $r_{i}^{+}$, and $n_{i}=1 \rightarrow 0$, with rate $r_{i}^{-}$. The rates may depend, in general, on the full set of states $r_{i}^{\pm}(\n)$, however, most common models assume a dependence only through the number of neighbors in state $1$, $q_{i}=\sum_{j=1}^{N} A_{ij} n_{j}$, in addition to the total number of neighbors $k_{i}$. For this reason we will term the individual rates as $r_{i}^{\pm}(\n) \equiv R_{k_{i},q_{i}}^{\pm}$, depending only on $k_{i},q_{i}$.

In our study we will focus on global quantities, such as the total density of nodes in state $1$, $m \equiv \frac{1}{N}\sum_{i=1}^{N} n_{i} \in (0, 1)$. For symmetrical models $R_{k,q}^{+}=R_{k,k-q}^{-}$ it is more natural to define the \emph{magnetization} as $m_{S} \equiv 2 m-1 \in (-1, 1)$ and we will use one quantity or another depending on the symmetries. The density of active links $\rho$, i.e. links connecting nodes in different states, is computed as
\begin{equation}
\label{density_active}
\rho \equiv \frac{\sum_{i,j=1}^{N} A_{ij} \left( n_{i}(1-n_{j})+(1-n_{i}) n_{j} \right)}{\sum_{i,j=1}^{N} A_{ij}}.
\end{equation}
One of the interesting properties of $\rho$ for binary-state models is that it can be used as an alternative to $m$ or $m_S$ to measure the level of order or agreement on one of the options, a situation in which $\rho$ approaches zero, independently of the option.

The stochastic dynamics produces variability across realizations/trajectories of the stochastic process. For this reason, one performs an average over realizations of the macroscopic quantities $\langle m(t) \rangle$, $\langle \rho(t) \rangle$ to characterize the global state of the system. A way to measure fluctuations and variability across realizations is by calculating the variance of the magnetization:
\begin{equation}
\label{susceptibility}
\chi \equiv N \left( \langle m^2 \rangle - \langle m \rangle^2 \right),
\end{equation}
which is also traditionally called magnetic susceptibility in spin models, as it also quantifies how the system responds to an external perturbation such as a magnetic field. Note that after the average over the ensemble of realizations/trajectories is produced, one usually performs additional averages over the ensemble of networks generated with the configuration model with the same degree distribution $P_{k}$. This is because we consider the degree distribution as the only relevant characteristic of the network.

\section{The master equation}\label{sec_master}

The most detailed characterization of models whose dynamics is defined by stochastic rules is achieved by the knowledge of the probability $P(\x,t)$ of finding the system in state $\x$ at time $t$. The time-evolution of this probability is governed by a {\slshape master equation}. In order to construct a general master equation we consider: (i) a set of integer variables $\x \equiv (x_1, \dots, x_{M})$, and (ii) a set of processes $\nu = 1, \dots, K$ characterized by the changes in the variables $x_{j} \rightarrow x_{j} + \ell_{j}^{(\nu)}$, $j=1,\dots,M$, with rates $W^{(\nu)}(\x)$. Once we have these ingredients the general master equation reads \cite{VanKampen:2007,Gardiner:2009,Toral:2014}:
\begin{equation}
\label{master_eq}
\dfrac{\partial P(\x;t)}{\partial t}=\sum_{\nu=1}^K\left(\prod_{j=1}^M E_j^{-\ell_j^{(\nu)}}-1\right)\left[W^{(\nu)}(\x)P(\x;t)\right],
\end{equation}
where $E_j$ is the step operator acting on any function $f(\x)$ of the variable $x_j$ as $E^\ell_j\left[f(x_1,\dots,x_j,\dots,x_M)\right]=f(x_1,\dots,x_j+\ell,\dots,x_M)$. For example, if we choose to include in our description the full set of node-state variables $\x=\n$, we have the following $K=2N$ processes: $\nu = (i,+)$ where $n_{i}=0 \rightarrow 1$, and $\nu=(i,-)$ where $n_{i}=1 \rightarrow 0$, for $i=1,\dots,N$. The changes in the variables are $\ell_{j}^{(i,\pm)}=\pm \delta_{i,j}$ and the respective rates $W^{(i,+)}= \delta_{n_{i},0} r_{i}^{+}$ and $W^{(i,-)}= \delta_{n_{i},1} r_{i}^{-}$ ($\delta_{n_{i},0} = 1 - n_{i}$, $\delta_{n_{i},1} = n_{i}$).

When the individual rates $r_i^\pm$ depend only on the number $k_i$ of neighbors and the number $q_i$ of those in the state $1$, $r_{i}^{\pm} = R_{k_{i},q_{i}}^{\pm}$ an alternative to the description based on the full set of node-state variables is to consider a compartmental approach also known as AME~\footnote{Note that in the acronym AME, or Approximate Master Equation, the word ``Master'' has a different meaning than the one used in this paper to describe an equation for the probability distribution.}. This mesoscopic description in terms of the number of nodes with the same transition rate, was studied in detail in \cite{Gleeson:2011,Gleeson:2013} and generalizations of this approach have been developed for multi-state models \cite{Fennell:2019} and weighted networks \cite{Unicomb:2018}. The {\slshape occupation numbers} are defined as the number of nodes $\x \equiv \lbrace N_{n, k, q} \rbrace$ that are in state $n = 0, 1$ and have degree $k = k_{\text{min}}, k_{\text{min}}+1, \dots, k_{\text{max}}$ among which $q = 0, 1, \dots, k$ are adopter neighbor nodes (nodes in state $1$). The level of description consists of $M=\sum_{k,q} 2 = (1+k_{\text{max}}-k_{\text{min}})(2+k_{\text{max}}+k_{\text{min}})$ variables, which are not all independent. The total number of nodes that have degree $k$ is fixed by the network, i.e. $N_{k} = \sum_{n,q} N_{n,k,q}$, which constitutes a total of $k_{\text{max}}-k_{\text{min}}+1$ constraints between variables. Another more subtle constraint is that in an undirected network the number of $0$-$1$ links is equal to the number of $1$-$0$, i.e. $\sum_{k,q} q N_{0,k,q} = \sum_{k,q} (k-q) N_{1,k,q}$. Interestingly, in the limit of uncorrelated networks $k_{\text{max}} \propto \sqrt{N}$ it is $M \propto N$, which indicates that the number of variables is of a similar magnitude compared to the node-state approach. Consequently, the occupation number approach will correspond to a significant decrease in the number of variables only when the degree distribution extends over a limited range of degree values $k_{\text{max}} \ll \sqrt{N}$. The global variables of interest, used to portray the macroscopic state of the system, are the total number of adopter nodes $N_{1} = \sum_{k,q} N_{1,k,q}$ and the number of active links (connecting nodes in state $0$ to $1$ or vice-versa) $L=\sum_{k,q} q N_{0,k,q}$, and their respective densities $m = N_{1}/N$, $\rho=2 L / (\mu N)$ defined in Section \ref{sec_general}.

\begin{figure*}[t]
\centering
\includegraphics[width=0.45\textwidth]{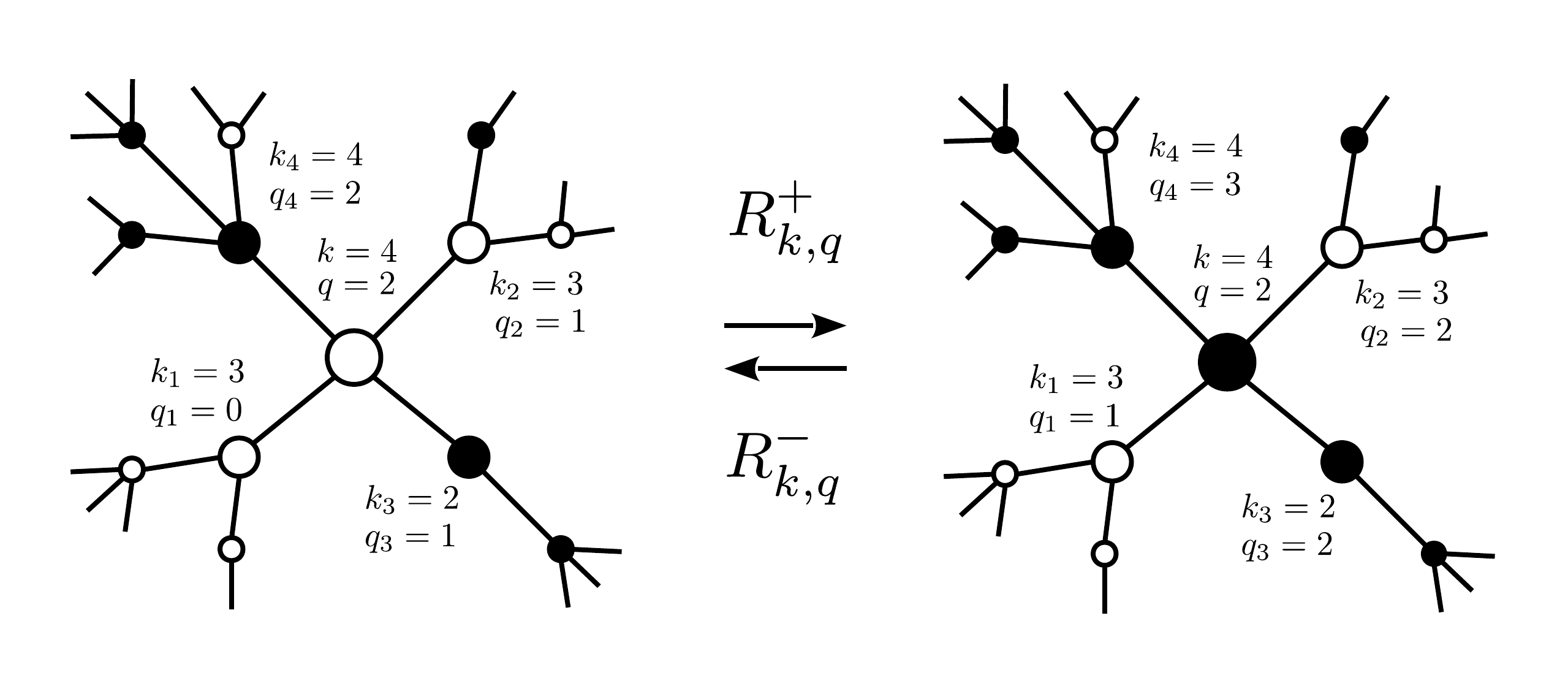}
\includegraphics[width=0.45\textwidth]{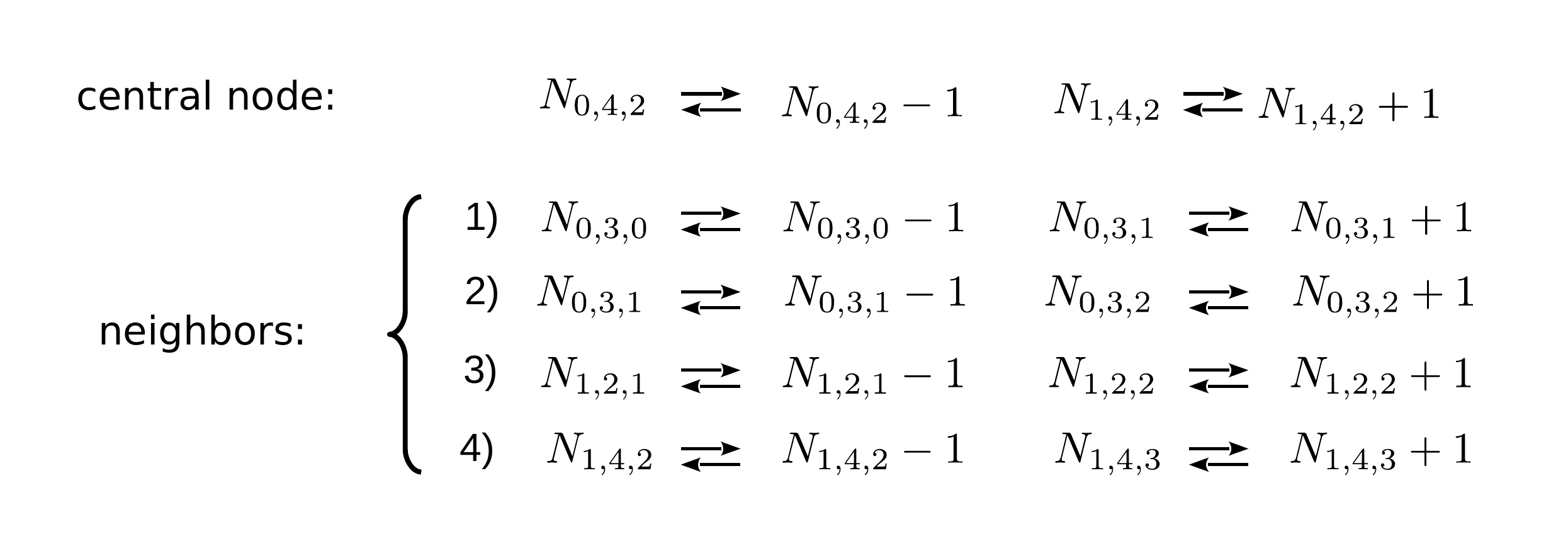}
\caption{Diagram of changes in the description variables $\lbrace N_{n, k, q} \rbrace$ when a node with $(k,q)=(4,2)$ changes state in a particular given network configuration, with neighborhood $(k_1,k_2,k_3,k_4) = (3,3,2,4)$ and $(q_1,q_2,q_3,q_4) = (0,1,1,2)\rightleftharpoons (1,2,2,3)$. Ten total changes are produced in the variables, two for the variables associated to the central node $k,q$ and two additional changes for each one associated to the neighbors $\lbrace k_{i},q_{i} \rbrace_{i=1, \dots, k}$.}
\label{fig:scheme}
\end{figure*}
In this occupation number approach $\{N_{n,k,q}\}$, however, the construction of the master equation is more cumbersome as we need to identify the possible processes $\nu$ and the associated rates $W^{(\nu)}$, and this will be our concern in the remainder of this section. Still, the elementary process of the dynamic is the state transition of a node $i$ compatible with the number $N_{n,k,q}$ changing from $n_{i}=0$ to $n_{i}=1$ or viceversa, but all processes that lead to the same change of the occupation number variables are grouped under the same label $\nu$. In an elementary process, $2(k+1)$ changes of the set of description variables $\{N_{n,k,q}\}$ are produced, two for the variables associated to the central node and two for each one of its neighbors, see Fig. \ref{fig:scheme} as a schematic example. The variables that change during this process depend on the values $k,q$ of the chosen node $i$, and additionally on the set $\lbrace k_{j},q_{j} \rbrace_{j=1, \dots,k}$ of the $k$ neighbors of $i$. We adopt to order the list of neighbors such that $\lbrace k_{j},q_{j} \rbrace_{j=1, \dots,k-q}$ correspond to the neighbors in state $0$, and $\lbrace k_{j},q_{j} \rbrace_{j=k-q+1, \dots, k}$ to the neighbors in state $1$. Therefore, the characterization of a process $\nu$ requires of the knowledge of the full set of variables, i.e. $\nu=(n,k,q,\lbrace k_{j} , q_{j} \rbrace_{j=1,\dots,k})$. The problem now is that, in principle, one is not able to know from the variables $\lbrace N_{n,k,q} \rbrace$ the set $\lbrace k_{j} , q_{j} \rbrace_{j=1,\dots,k}$, and consequently we need some approximation to attain a closed treatment of the dynamics. We make the ansatz that the rate of each process is calculated as the total change rate of the central node $N_{n,k,q} R_{k,q}^{\pm}$ times the probability of having a particular configuration of the neighborhood, this is:
\begin{widetext}
\begin{eqnarray}
\label{rates1}
W^{(0,k,q,\lbrace k_{j},q_{j} \rbrace_{j=1, \dots,k})} \left(\{N_{n,k,q}\}\right)&=& N_{0,k,q} R_{k,q}^{+} \prod_{j=1}^{k-q} P_{0}(0,k_{j},q_{i}) \prod_{j=k-q+1}^{k} P_{0}(1,k_{j},q_{j}),\\
\label{rates2}
W^{(1,k,q,\lbrace k_{j},q_{j} \rbrace_{j=1, \dots,k})}\left(\{N_{n,k,q}\}\right) &=& N_{1,k,q} R_{k,q}^{-} \prod_{i=1}^{k-q} P_{1}(0,k_{i},q_{i}) \prod_{j=k-q+1}^{k} P_{1}(1,k_{j},q_{j}).
\end{eqnarray}
\end{widetext}
Here, we introduced $P_{0}(1,k_{j},q_{j})$, defined as the probability that an edge leaving a node in state $0$ connects to a node in state $1$ with $k_{j},q_{j}$, and equivalently for $P_{0}(0,k_{i},q_{i})$, $P_{1}(0,k_{i},q_{i})$ and $P_{1}(1,k_{j},q_{j})$. These probabilities can be calculated using the description variables $N_{n,k,q}$ as:
\begin{eqnarray}
\label{probability00}
P_{0}(0,k,q)&=&\frac{(k-q)N_{0,k,q} }{\sum_{k,q} (k-q)N_{0,k,q} }, \\
\label{probability01}
P_{0}(1,k,q)&=&\frac{(k-q)N_{1,k,q} }{\sum_{k,q} (k-q)N_{1,k,q} }, \\
\label{probability10}
P_{1}(0,k,q)&=&\frac{q N_{0,k,q} }{\sum_{k,q} q N_{0,k,q}}, \\
\label{probability11}
P_{1}(1,k,q)&=&\frac{q N_{1,k,q}}{\sum_{k,q} q N_{1,k,q}}.
\end{eqnarray}
For example, $P_{0}(1,k,q)$ is the fraction of edges coming out of nodes in state $1$ with $k, q$ that go to nodes in state $0$, divided by the total number of $0$-$1$ edges and similarly for the other expressions. Note that the approximation in this method is that we assume the probability of the neighborhood configuration of a node to be a product of independent single event probabilities, which is of general validity for uncorrelated networks.

We now define $\ell_{n,k,q}^{(\nu)}$ as the change in the variable $N_{n,k,q} \rightarrow N_{n,k,q}+\ell_{n,k,q}^{(\nu)}$ in the process $\nu$, which are computed as (see Fig. \ref{fig:scheme} for a guide):
\begin{widetext}
\begin{eqnarray}
\label{changes1}
 \ell_{0,k,q}^{(0,k',q',\lbrace k_{j},q_{j} \rbrace_{j=1,\dots,k})} &=& -\delta_{k,k'} \delta_{q,q'} + \sum_{j=1}^{k'-q'} \left( - \delta_{k,k_{j}} \delta_{q,q_{j}} + \delta_{k,k_{j}} \delta_{q,q_{j}+1}\right), \\
 \label{changes2}
 \ell_{1,k,q}^{(0,k',q',\lbrace k_{j},q_{j} \rbrace_{j=1,\dots,k})} &=& \delta_{k,k'} \delta_{q,q'} + \sum_{j=k'-q'+1}^{k'} \left( - \delta_{k,k_{j}} \delta_{q,q_{j}} + \delta_{k,k_{j}} \delta_{q,q_{j}+1}\right), \\
 \label{changes3}
 \ell_{0,k,q}^{(1,k',q',\lbrace k_{j},q_{j} \rbrace_{j=1,\dots,k})} &=& \delta_{k,k'} \delta_{q,q'} + \sum_{j=1}^{k'-q'} \left( - \delta_{k,k_{j}} \delta_{q,q_{j}} + \delta_{k,k_{j}} \delta_{q,q_{j}-1}\right), \\
 \label{changes4}
 \ell_{1,k,q}^{(1,k',q',\lbrace k_{j},q_{j} \rbrace_{j=1,\dots,k})} &=& -\delta_{k,k'} \delta_{q,q'} + \sum_{j=k'-q'+1}^{k'} \left( - \delta_{k,k_{j}} \delta_{q,q_{j}} + \delta_{k,k_{j}} \delta_{q,q_{j}-1}\right).
\end{eqnarray}
\end{widetext}
Once the processes $\nu$, rates $W^{(\nu)}$ and changes in the variables $\ell_{n,k,q}^{(\nu)}$ are defined, we can draw on the general theory of stochastic processes \cite{VanKampen:2007, Gardiner:2009, Toral:2014,Peralta_moments:2018} in terms of the master equation (\ref{master_eq}). 

Coarser levels of description are also possible. Let $N_{n,k}=\sum_qN_{n,k,q}$ be the number of nodes in state $n$ with degree $k$, and $L_{n,k}=\sum_{q} q N_{n,k,q}$ the number of links that connect nodes of degree $k$ and state $n$ with nodes in state $1$ (adopter nodes). The next level of description is the Pair Approximation (PA) that considers the set ${\x}=\{N_{1,k},L_{0,k},L_{1,k}\}$, with $N_{0,k}=N_{k}-N_{1,k}$, and $k\in[k_\text{min},k_\text{max}]$. The pair approximation reduces the number of variables to $M=3 (k_{\text{max}}-k_{\text{min}}+1)$ with the conservation of the total number of $0$-$1$ links $\sum_{k} L_{0,k} = \sum_{k} \left( N_{1,k} k - L_{1,k} \right)$ as the only constraint. The master equation requires to write the rates $W^{(\nu)}$ as a function only of the description variables. To achieve this, one introduces an approximation based on the ansatz that the variable $N_{n,k,q}$ appearing in the rates Eqs. (\ref{rates1}, \ref{rates2}) can be expressed as
\begin{eqnarray}
\label{pair_approximation}
N_{n,k,q} = N_{n,k} \text{Bin}_{k,q} \left[ \frac{L_{n,k}}{k N_{n,k}} \right],
\end{eqnarray}
where $\text{Bin}_{k,q} [p] = \binom{k}{q} p^q (1-p)^{k-q}$ is the binomial distribution. In this paper, we restrict our study to this version of the pair approximation, but other variants exist in the literature, such as the so-called heterogeneous pair approximation \cite{Pugliese:2009}, where one includes in the description the number of active links $L_ {k,k'}$ that join nodes of degree $k$ and $k'$ that are in different states, or the original version \cite{Vazquez:2008} (also called homogeneous pair approximation) that takes into consideration just the global density $L$ of active links. 

An even cruder level of description is the Heterogeneous Mean Field (HMF), which considers the set of variables $\x = \lbrace N_{1,k} \rbrace$, with $N_{0,k}=N_{k}-N_{1,k}$, reducing the number of variables to $M=k_{\text{max}}-k_{\text{min}}+1$ with no constraints. The closure of the rates $W^{(\nu)}$ in terms of this set of variables is achieved by a similar binomial ansatz but with a simpler single event probability:
\begin{eqnarray}
\label{HMF_approximation}
N_{n,k,q} = N_{n,k} \text{Bin}_{k,q} \left[ \frac{\sum_{k} k N_{1,k} }{\mu N } \right].
\end{eqnarray}

The coarsest possible description is the Mean Field (MF) in which a single description variable $\x=N_1$ is used with closure ansatz $N_{1,k,q}=N_1 \delta_{q,k N_{1}/N},\,N_{0,k,q}=(N-N_1)\delta_{q,k N_{1}/N}$.

Note that while the formulation of the node-state approach does not need any approximations, the different occupation number approaches, whether AME, PA, HME or MF, use approximations in the calculation of the rates that limit the validity of their predictions. In particular, as discussed previously, we expect the AME to be accurate only for uncorrelated networks. The fact that the master equation for the node-state approach is free of approximations does not mean in general that we are able to solve such equation, and different approximations are then needed to obtain a solution \cite{Lafuerza:2013,Carro:2016}. The advantage of the occupation number approach is that it has some particularities that enables us to apply accurate methods to solve the master equation. These different techniques are explained and explored in the next sections.

\section{Approximate solution of the master equation}

\subsection*{Formulation}
The main reason of the convenience of the occupation number approach is that the description variables are extensive. This means that for a fixed degree distribution $P_{k}$, if we increase the system size $N \rightarrow \lambda N$, the variables scale in the same way $N_{n,k,q} \rightarrow \lambda N_{n,k,q}$ and similarly for $N_{n,k}$, $L_{n,k}$, $N_{1,k}$ and $N_1$. This property is useful because it allows us to apply the well known system-size expansions of the master equation. Note that the rates Eqs. (\ref{rates1},\ref{rates2}) are extensive functions $W^{(\nu)}(\x) = N w^{(\nu)} \left( \frac{\x}{N} \right)$, where $N=\sum_{n,k,q} N_{n,k,q}$ is the total number of nodes and $w^{(\nu)}$ are the set of intensive rate functions. In this case, following \cite{Peralta_moments:2018, Peralta_pair:2018}, we can use a van Kampen type of system-size expansion, that we now explain in detail.

In the case of the AME, the expansion splits the variables as $\x=N \bphi+N^{1/2}\pmb{a}+N^{0}\pmb{b}$, in components $N_{n,k,q} = N \phi_{n,k,q} + N^{1/2} a_{n,k,q} + N^{0} b_{n,k,q}$, where $\phi_{n,k,q}$ are a set of deterministic variables, while $a_{n,k,q}$ and $b_{n,k,q}$ are random variables. This is an expansion which is assumed to be of general validity in the thermodynamic limit $N \rightarrow \infty$ and which yields the first stochastic correction terms to the deterministic approach \cite{Gleeson:2011, Gleeson:2013}. The deterministic evolution of the system fulfills a set of nonlinear differential equations 
\begin{equation}
\label{dphidt}
\frac{d\phi_{n,k,q}}{dt} = \Phi_{n,k,q},
\end{equation}
characterized by the drift term defined as $\Phi_{n,k,q}(\bphi) = \sum_{\nu} \ell_{n,k,q}^{(\nu)}w^{(\nu)}(\bphi)$ which leads, after some algebra using Eqs. (\ref{rates1}-\ref{changes4}), to:
\begin{eqnarray}
\label{Phi0}
\Phi_{0,k,q} &=& -\phi_{0,k,q} R_{k,q}^{+} + \phi_{1,k,q} R_{k,q}^{-} - \phi_{0,k,q} (k-q) \beta^{s} \notag\\
 &&+ \phi_{0,k,q-1} (k-q+1) \beta^{s} -\phi_{0,k,q} q \gamma^{s} \notag\\ && +\phi_{0,k,q+1} (q+1) \gamma^{s}, \\
 \label{Phi1}
 \Phi_{1,k,q} &=& \phi_{0,k,q} R_{k,q}^{+} - \phi_{1,k,q} R_{k,q}^{-} - \phi_{1,k,q} (k-q) \beta^{i} \notag\\
 &&+ \phi_{1,k,q-1} (k-q+1) \beta^{i}- \phi_{1,k,q} q \gamma^{i} \notag \\ &&+ \phi_{1,k,q+1} (q+1) \gamma^{i}.
\end{eqnarray}
Here $\beta^{s}$, $\gamma^{s}$, $\beta^{i}$ and $\gamma^{i}$ are the individual rates $R_{k,q}^{\pm}$ at which a neighbor of a central node changes state averaged with the probabilities Eqs. (\ref{probability00}-\ref{probability11}), where the symbol $\beta, \gamma$ reflects the state of the neighbor node $0, 1$, while the super index $s,i$ reflects the state of the central node $0,1$, namely:
\begin{align}
\label{betas}
\beta^{s} &\equiv \sum_{k,q} P_{0}(0,k,q) R_{k,q}^{+} = \frac{\sum_{k,q} (k-q)\,\phi_{0,k,q} R_{k,q}^{+}}{\sum_{k,q} (k-q)\,\phi_{0,k,q} },\\
\label{gammas}
\gamma^{s} &\equiv \sum_{k,q} P_{0}(1,k,q) R_{k,q}^{-} = \frac{\sum_{k,q} (k-q)\,\phi_{1,k,q} R_{k,q}^{-}}{\sum_{k,q} (k-q)\,\phi_{1,k,q} },\\
\label{betai}
\beta^{i} &\equiv \sum_{k,q} P_{1}(0,k,q) R_{k,q}^{+} = \frac{\sum_{k,q} q\,\phi_{0,k,q} R_{k,q}^{+}}{\sum_{k,q} q\, \phi_{0,k,q}}, \\ 
\label{gammai}
\gamma^{i} &\equiv \sum_{k,q} P_{1}(1,k,q) R_{k,q}^{-} = \frac{\sum_{k,q} q\, \phi_{1,k,q} R_{k,q}^{-}}{\sum_{k,q}q\, \phi_{1,k,q}}.
\end{align}
Note that, at the deterministic level, the set of differential equations (\ref{dphidt}-\ref{gammai}) coincides with the original work of Gleeson \cite{Gleeson:2011,Gleeson:2013}, as it is naturally expected. The advantage of the stochastic formalism presented here Eqs. (\ref{master_eq}-\ref{changes4}) is that, in addition, we will be able to obtain results for the average deviations $\langle a_{n,k,q} \rangle$, $\langle b_{n,k,q} \rangle$ from the deterministic solution, and also for the fluctuations/correlations $C_{n,k,q;n',k',q'}=\langle a_{n,k,q} a_{n',k',q'} \rangle-\langle a_{n,k,q} \rangle \langle a_{n',k',q'} \rangle$. In the van Kampen expansion, the set of differential equations for these quantities are linear and read in vector notation \cite{Peralta_moments:2018}:
\begin{eqnarray}
\label{average_a}
\frac{d\langle \pmb{a} \rangle}{dt} &=& - \mathbf{B} \langle \pmb{a} \rangle, \\
\label{average_b}
\frac{d\langle \pmb{b} \rangle}{dt} &=& - \mathbf{B} \langle \pmb{b} \rangle + \mathbf{\Gamma}, \\
\label{corr_aa}
\frac{d \mathbf{C}}{dt} &=& - \mathbf{B} \mathbf{C} - \mathbf{C} \mathbf{B} + \mathbf{G},
\end{eqnarray}
where $\mathbf{B}$ is the Jacobian matrix $B_{ij}(\bphi) = - \partial_{\phi_{j}} \Phi_{i}$; the noise $\mathbf{G}$ matrix is calculated as $G_{ij}(\bphi) = \sum_{\nu} \ell_{i}^{(\nu)} \ell_{j}^{(\nu)} w^{(\nu)}(\bphi)$ and $\Gamma_{i}= \frac{1}{2} \sum_{j,k} \langle a_{j} a_{k} \rangle \partial^2_{\phi_{j},\phi_{k}} \Phi_{i}$ is related to the Hessian matrices of $\pmb{\Phi}$. For reason of space, the explicit expressions of these matrices are written down in Appendix \ref{app:matrices}. 

In the case of the Pair Approximation, and proceeding with the general theory, we split the variables like $N_{1,k}=N \phi_{k} + N^{1/2} a_{k} + N^{0} b_{k}$ and $L_{n,k}= N \phi_{n,k} + N^{1/2} a_{n,k} + N^{0} b_{n,k}$. The evolution equation at the deterministic level is:
\begin{eqnarray}
\frac{d\phi_{k}}{dt}&=&\Phi_{k},\\
\frac{d\phi_{n,k}}{dt}&=&\Phi_{n,k}.
\end{eqnarray}
In order to obtain the deterministic drift functions $\pmb{\Phi}$, we have to perform sums in Eqs. (\ref{Phi0}-\ref{Phi1}) as $\Phi_{k} = \sum_{q} \Phi_{1,k,q}$ and $\Phi_{n,k} = \sum_{q} q \Phi_{n,k,q}$, which leads to
\begin{eqnarray}
\label{PhiPA1}
\Phi_{k} &=& \sum_{q} \left[ \phi_{0,k,q} R_{k,q}^{+} - \phi_{1,k,q} R_{k,q}^{-} \right], \\
\label{PhiPA2}
\Phi_{0,k} &=& \sum_{q} \left[ -q \phi_{0,k,q} R_{k,q}^{+} + q \phi_{1,k,q} R_{k,q}^{-} \right] \notag\\ &&+ \beta^{s} (k P_{k} - k \phi_{k} - \phi_{0,k}) - \gamma^{s} \phi_{0,k},\\
\Phi_{1,k} &=& \sum_{q} \left[ q \phi_{0,k,q} R_{k,q}^{+} - q \phi_{1,k,q} R_{k,q}^{-} \right]  \notag\\ &&+ \beta^{i} (k \phi_{k} -\phi_{1,k}) - \gamma^{i} \phi_{1,k},
\end{eqnarray}
where one must replace $\phi_{n,k,q}$ by the binomial ansatz $\phi_{0,k,q} = (P_k-\phi_k) \text{Bin}_{k,q} \left[ p_{0,k} \right]$, $\phi_{1,k,q} = \phi_k \text{Bin}_{k,q} \left[ p_{1,k} \right]$ and $p_{0,k}=\phi_{0,k}/(k (P_k-\phi_k))$, $p_{1,k}=\phi_{1,k}/(k \phi_k)$. The corresponding Jacobian $\mathbf{B}$ and $\mathbf{G}$ matrices of this Pair Approximation can be found in Appendix \ref{app:matrices}.

In the case of the Heterogeneous Mean Field, the variable splitting in this case is $N_{1,k}= N \phi_{k} + N^{1/2} a_{k} + N^{0} b_{k}$, and deterministic equation $\dfrac{d\phi_k}{dt}=\Phi_k$, where the drift functions $\pmb{\Phi}$ are obtained by summing Eqs. (\ref{Phi0}-\ref{Phi1}) like $\Phi_{k} = \sum_{q} \Phi_{1,k,q}$, which leads also to
\begin{equation}
\label{Phi_HMG}
\Phi_{k} = \sum_{q} \left[ \phi_{0,k,q} R_{k,q}^{+} - \phi_{1,k,q} R_{k,q}^{-} \right],
\end{equation}
but now $\phi_{n,k,q}$ are given by $\phi_{0,k,q} = (P_k-\phi_k) \text{Bin}_{k,q} \left[ p \right]$, $\phi_{1,k,q} = \phi_k \text{Bin}_{k,q} \left[ p \right]$ with $p=\sum_{k}k\phi_{k} /\mu$ (independent of $k$). Again, the corresponding Jacobian $\mathbf{B}$ and $\mathbf{G}$ matrices can be found in Appendix \ref{app:matrices}.

In a previous work \cite{Peralta_moments:2018} we explained how to solve equations (\ref{average_a}-\ref{corr_aa}) and we developed a very stable and fast convergent implicit Euler method to find the numerical solution of the correlation matrix $\mathbf{C}$. It is worth mentioning that a general result of the van Kampen expansion is that the stationary probability distribution $\Pi_{\text{st}}(\pmb{a})$ of the first stochastic correction $\pmb{a}$ is Gaussian \cite{VanKampen:2007,Peralta_moments:2018} with zero mean $\langle \pmb{a} \rangle_{\text{st}}=0$:
\begin{equation}
\label{Gaussian}
\Pi_\text{st}(\pmb{a})=\sqrt{\frac{|{\mathbf C_\text{st}}|}{(2\pi)^M}}e^{-\frac12 \pmb{a}^\intercal \cdot {\mathbf{C}_\text{st}^{-1}} \cdot \pmb{a}}.
\end{equation}
Besides, if the initial condition $\Pi(\pmb{a},t=0)$ is Gaussian, then the time-dependent $\Pi(\pmb{a},t)$ is also a Gaussian (\ref{Gaussian}) replacing the stationary correlation matrix ${\mathbf C_\text{st}} \rightarrow {\mathbf C(t)}$ by the time-dependent one.

The van Kampen expansion will be accurate in the thermodynamic limit $N \rightarrow \infty$, for example in the determination of the magnetic susceptibility, this is:
\begin{equation}
\label{susceptibility_vK}
\chi = \sum_{k,q,k',q'} C_{1,k,q;1,k',q'}.
\end{equation}
According to the van Kampen approach, the susceptibility defined as Eq. (\ref{susceptibility}) does not depend on system size $N$, which is obviously not true for a finite system $N$. What we are obtaining in this approach is the thermodynamic limit $\lim_{N \rightarrow \infty} \chi_{N}$. With respect to the average values of the macroscopic quantities $\langle m(t) \rangle$, $\langle \rho(t) \rangle$ they are computed as:
\begin{eqnarray}
\label{average_values}
\langle m(t) \rangle &=& \sum_{k,q} \phi_{1,k,q} + \frac{1}{N}\sum_{k,q} \langle b_{1,k,q} \rangle, \\ 
\langle \rho(t) \rangle &=& \sum_{k,q} q \phi_{0,k,q} + \frac{1}{N}\sum_{k,q} q \langle b_{0,k,q} \rangle.
\end{eqnarray}
This is nothing but the deterministic solution plus a correcting factor of order $O(N^{-1})$ (note that $\langle \pmb{a} \rangle = 0$ Eq. (\ref{average_a})).

An alternative less restrictive system size expansion is the Kramers-Moyal expansion, which transforms the master equation (\ref{master_eq}) into a continuous PDE for the intensive variables $\pmb{\varphi}$. If we define the densities $\pmb{\varphi} = \x/N$, the Kramers-Moyal expansion \cite{VanKampen:2007} leads to the Fokker-Planck equation \cite{SanMiguel:2000} for the probability density $\Pi(\pmb{\varphi};t)$ of the intensive variables:
\begin{align}
\label{FokkerPlanck_equation}
& \frac{\partial \Pi(\pmb{\varphi};t)}{\partial t} = \notag\\
& - \sum_{i=1}^{K} \frac{\partial}{\partial \varphi_{i}} \left[ - \Phi_{i}(\pmb{\varphi}) \Pi + \frac{1}{2N} \sum_{j=1}^{K} \frac{\partial}{\partial \varphi_{j}} \Big[ G_{ij}(\pmb{\varphi}) \Pi \Big] \right],
\end{align}
where $\Phi_{i}$ and $G_{ij}$ are the same functions defined previously. The problem with the Kramers-Moyal expansion is that, in most occasions, it is as complicated to solve as the original master equation (\ref{master_eq}), while the van Kampen expansion corresponds to a linearization of Eq. (\ref{FokkerPlanck_equation}) (this is why it is also called linear noise approximation) where we assume $\pmb{\varphi}$ to weakly fluctuate around the deterministic value $\pmb{\phi}$, this is $\pmb{\varphi} \approx \pmb{\phi} + N^{-1/2}\pmb{a}$.

In the next subsection we apply the van Kampen expansion method to several models of interest and check its accuracy and validity.

\subsection*{Comparison with numerical simulations}\label{sec_vKresults}
We will now compare the results of the theory explained in the previous section to the numerical simulations. We will focus on stationary quantities in order to study how the results change with the parameters of the models.

\begin{figure}[h!]
\includegraphics[width=0.45\textwidth]{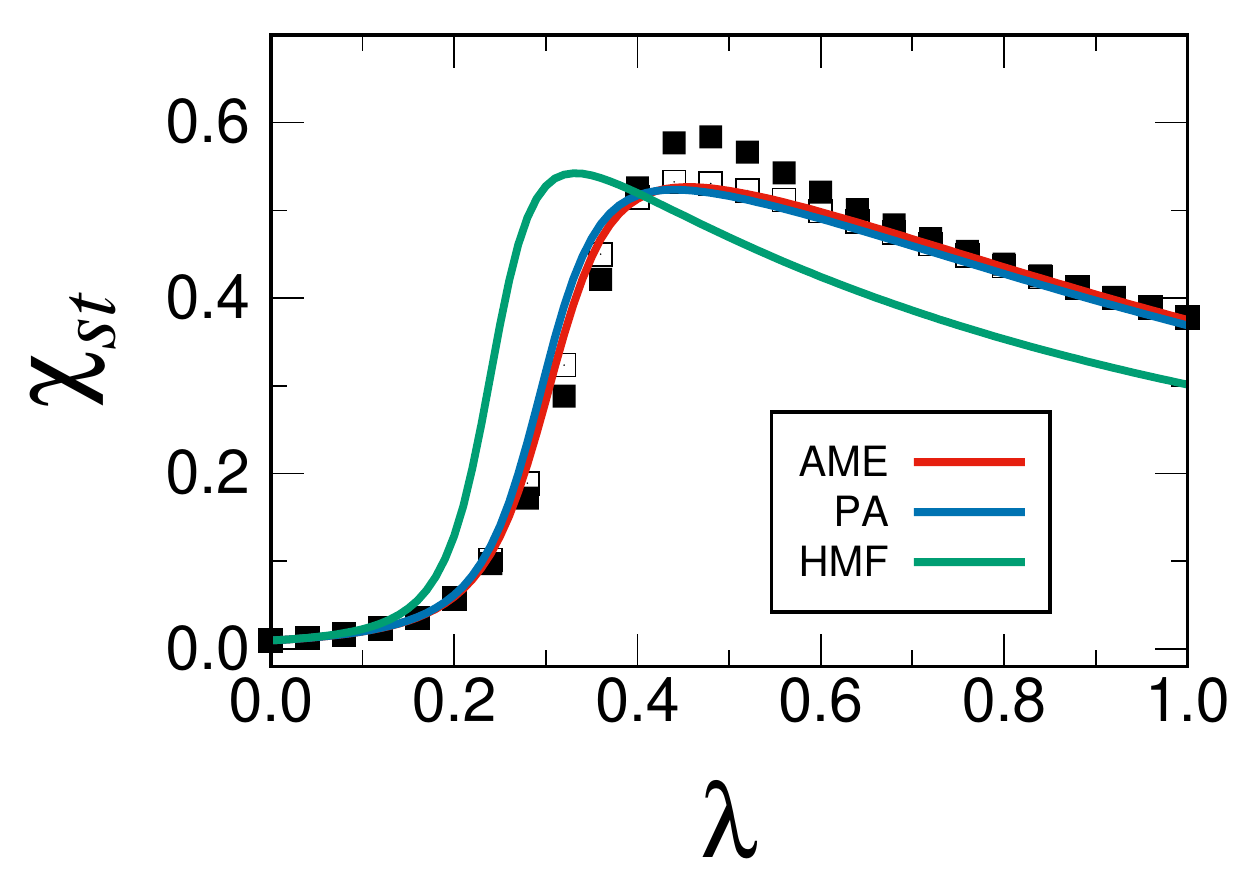}
\includegraphics[width=0.45\textwidth]{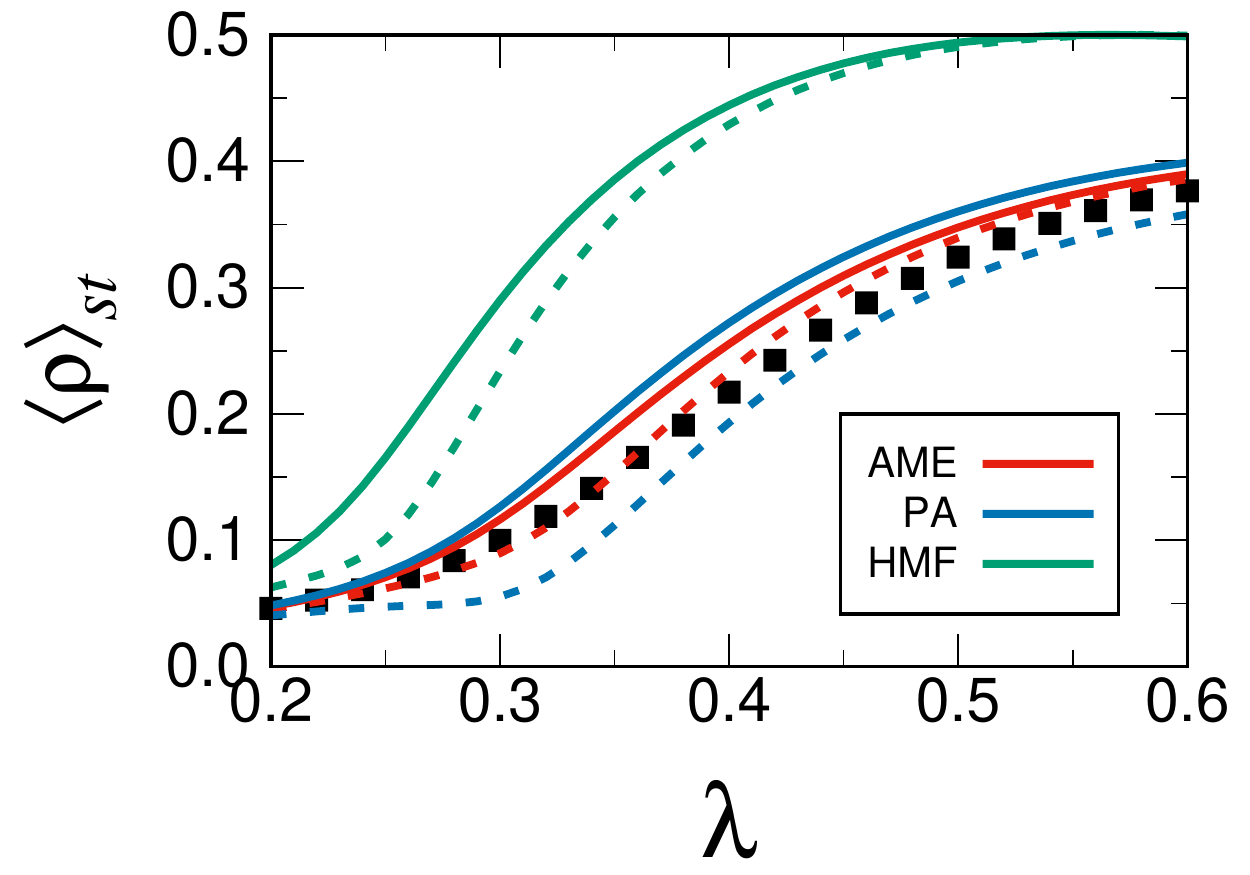}
\caption{Stationary susceptibility $\chi_{\text{st}}$ and average density of active links $\langle \rho \rangle_{\text{st}}$ as a function of the transmission rate $\lambda$ for the SIS model. We choose as parameters $\varepsilon=10^{-2}$ and $\mu=1$ on a scale free network $P_{k} \sim k^{-2.5}$ with $k_{\text{min}}=2$ and $k_{\text{max}}=10$. Points correspond to numerical simulations of the model with $N=100$ (solid squares) and $N=400$ (empty squares) averaged over an ensemble of $100$ networks. Lines of different colors are the theoretical prediction of the different approximations. In the left panel the solid lines are the van Kampen result Eq. (\ref{susceptibility_vK}), while in the right panel the solid line is the deterministic approach and the dashed lines the corrected average values Eq. (\ref{average_values}).}
\label{fig:SIS}
\end{figure}

The first model that we consider is the SIS (susceptible-infected-susceptible) epidemic model \cite{Bailey:1975,Anderson:1991} on a scale-free network with rates $R_{k,q}^{+} = \varepsilon + \lambda q$ and $R_{k,q}^{-}=\mu$. Here $\lambda$ is the transmission rate, $\mu$ is the recovery rate, and $\varepsilon$ is the rate at which an outbreak appears in the system. Note that we incorporate the parameter $\varepsilon$ in order for the system to have a properly defined stationary result, in principle one recovers the traditional SIS model by letting $\varepsilon \rightarrow 0$. In Fig. \ref{fig:SIS} we see that the van Kampen approach predicts accurately the stationary susceptibility for a small system size $N=100$ with increasing accuracy as it increases to $N=400$. The finite-size corrections to the average value $\langle \rho\rangle_\text{st}$ are plotted in Fig. \ref{fig:SIS}, with an improvement in the deterministic solution. If we focus on the comparison between the different approximations we observe, as expected, an increase in accuracy as $\text{AME} > \text{PA} > \text{HMF}$. Although the difference between the AME and the PA is small with respect to the deterministic $\pmb{\phi}$ and fluctuations $\chi$, the finite size corrections of the average values $N^{-1} \langle \pmb{b} \rangle$ are well captured only by the AME. This indicates that the PA is a good approximation at the deterministic and linear (Jacobian) level Eq. (\ref{average_a}, \ref{corr_aa}) but not for the second order correction (Hessian) level Eq. (\ref{average_b}).

\begin{figure}[h!]\includegraphics[width=0.45\textwidth]{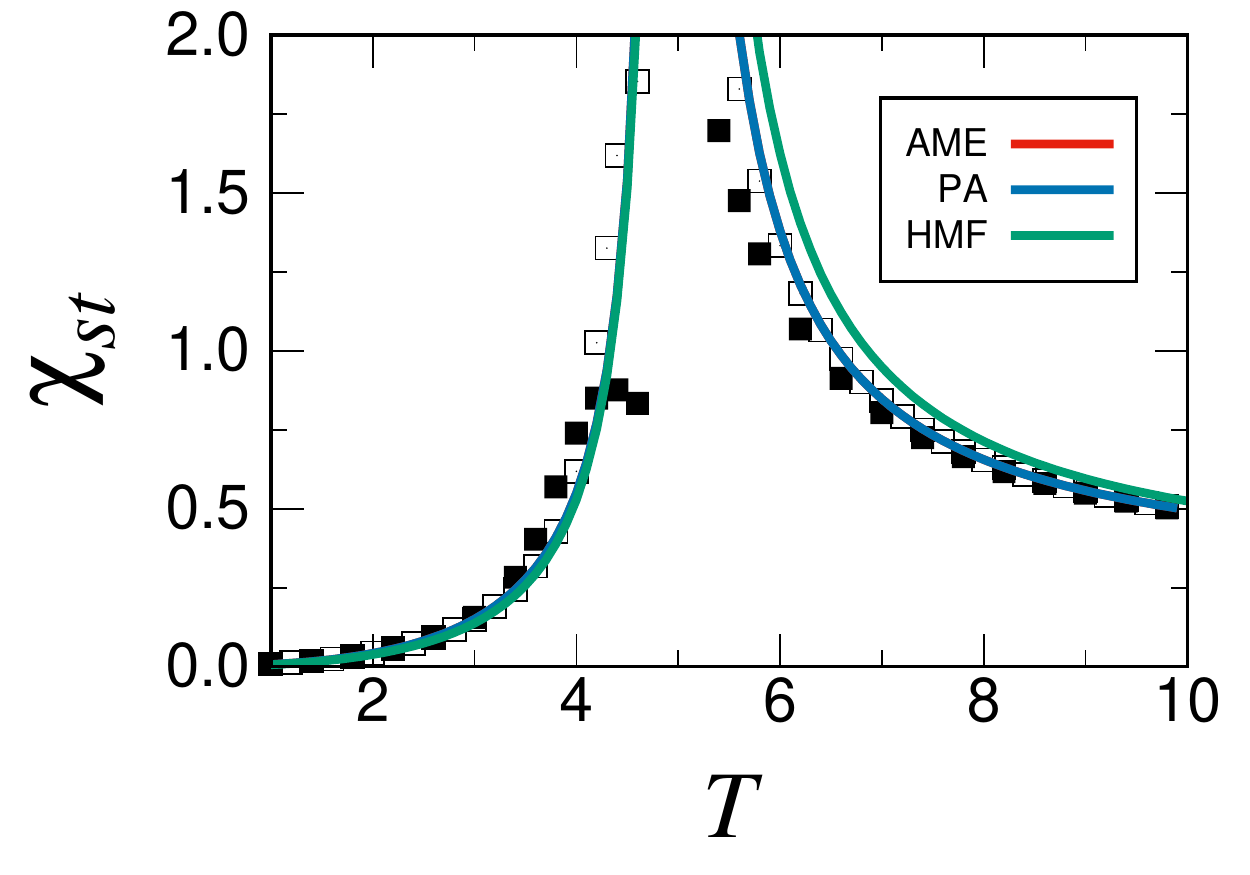}
\includegraphics[width=0.45\textwidth]{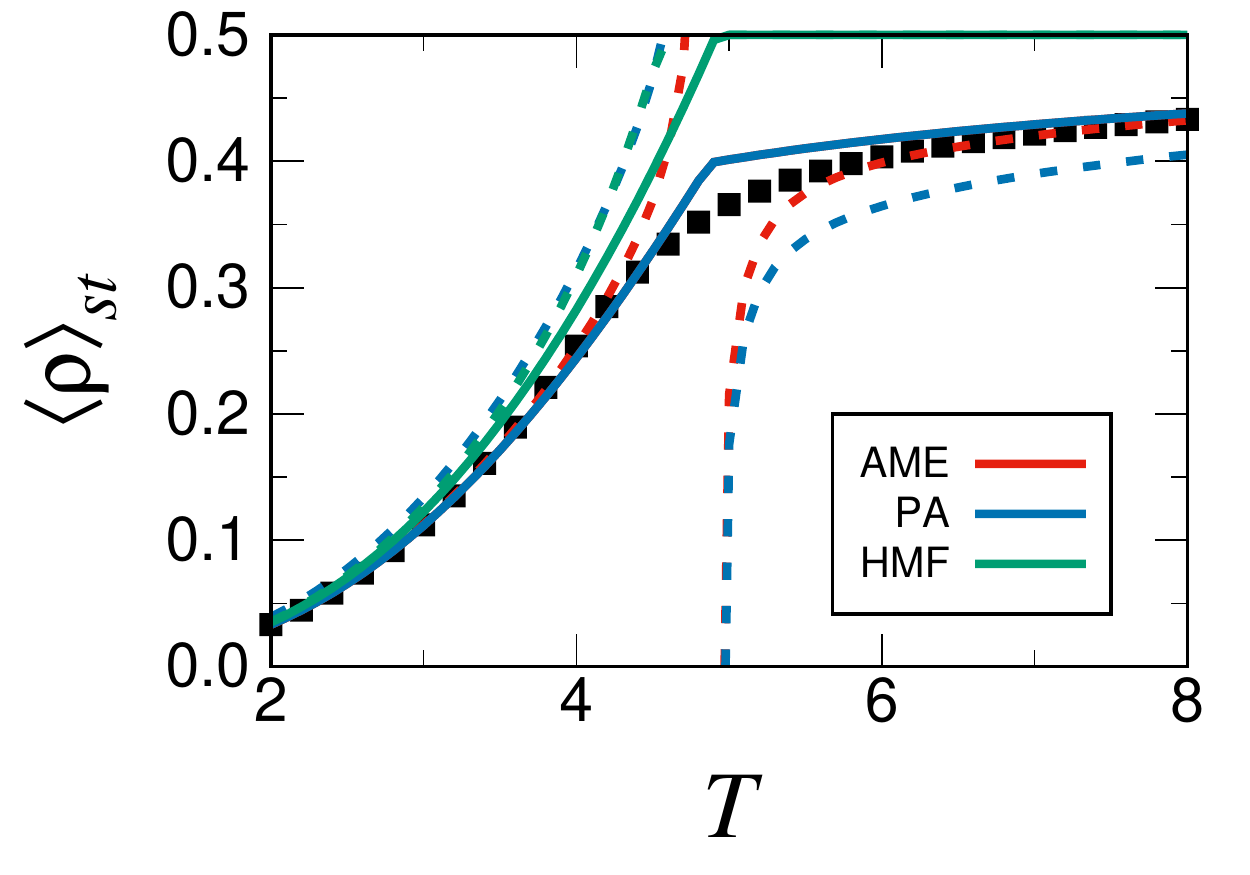}
\caption{Stationary susceptibility $\chi_{\text{st}}$ and average density of active links $\langle \rho \rangle_{\text{st}}$ as a function of the temperature $T$ for the Ising model with Glauber rates. We choose as parameter $J=1$ on an Erd\H{o}s-R\'enyi network with average degree $\mu=5$. Points correspond to numerical simulations of the model with $N=100$ (solid squares) and $N=400$ (empty squares) averaged over an ensemble of $100$ networks. Lines of different colors are the theoretical prediction of the different approximations (the curves that do not appear are superposed). In the left panel the solid lines are the van Kampen result Eq. (\ref{susceptibility_vK}), while in the right panel the solid line is the deterministic approach and the dashed lines the corrected average values Eq. (\ref{average_values}).}
\label{fig:Glauber}
\end{figure}

The second model to which we apply our theory is the Ising model defined on an Erd\H{o}s-R\'enyi network with Glauber rates \cite{Glauber:1963} $R_{k,q}^{+}=\left(1+e^{\frac{2J}{T}(k-2q)}\right)^{-1}$ and $R_{k,q}^{-}=R_{k,k-q}^{+}$, where $J$ is the coupling strength and $T$ the temperature. Note that it is a symmetric model and thus, as discussed in Section \ref{sec_general}, we choose $m_{S} = 2 m - 1$ to calculate the susceptibility. It is well known that the Ising model has a critical point $T_{c}$ such that $\lim_{N\rightarrow \infty} \langle \vert m_{S} \vert \rangle_{\text{st}}=0$ if $T > T_{c}$ and $\lim_{N\rightarrow \infty} \langle \vert m_{S} \vert \rangle_{\text{st}} \propto (T_{c}-T)^{\beta}$ if $T < T_{c}$, while $\langle m_{S} \rangle_{\text{st}}=0$ always for symmetry. For this reason, following the standard procedure, we compute the susceptibility numerically as $\chi_{\text{st}} = N \left( \langle m_{S}^2 \rangle_{\text{st}} - \langle \vert m_{S} \vert \rangle_{\text{st}}^2 \right)$ if $T < T_{c}$, and $\chi_{\text{st}} = N \langle m_{S}^2 \rangle_{\text{st}}$ if $T > T_{c}$. In principle we can not know the position of the critical point numerically for a single finite system size, we thus plot both quantities $\chi_{\text{st}} = N \left( \langle m_{S}^2 \rangle_{\text{st}} - \langle \vert m_{S} \vert \rangle_{\text{st}}^2 \right)$ and $\chi_{\text{st}} = N \langle m_{S}^2 \rangle_{\text{st}}$ and eliminate those points to the right of the peak of the first expresion. In Fig. \ref{fig:Glauber} we see a good prediction of the stationary susceptibility for the small system $N=100$ with increasing accuracy for $N=400$. Note that the theory predicts the divergence of the susceptibility at the critical point $T_c$, as $\chi_{\text{st}} \propto \vert T-T_{c} \vert^{-\gamma}$, which can be strictly true only in the thermodynamic limit $N \rightarrow \infty$. For a finite system it can be shown, see Section \ref{sec_critical}, that if we approach the critical point as an inverse power of the system size $\vert T-T_{c} \vert \propto N^{-r}$, the susceptibility scales as a positive power $\chi_{\text{st}} \propto N^{2 \upsilon - 1}$, with appropriate exponents $r > 0$ and $\upsilon > 1/2$ determined in Section \ref{sec_critical}. This implies that the van Kampen expansion presents discrepancies with the numerical results that are important, for a finite system $N$, in a region of the critical point whose width decreases with system size. Similarly, we see in Fig. \ref{fig:Glauber} that the stochastic correction to the density of active links $\langle\rho\rangle_\text{st}$ is only accurate outside the critical region, while it diverges at the critical point, at odds with the numerical result. In the comparison between the different approximations we observe again an increase in accuracy as $\text{AME} > \text{PA} > \text{HMF}$. As proven in \cite{Gleeson:2013} the deterministic part of the PA and AME are completely equivalent for all models fulfilling the microscopic reversibility condition $R_{k,q}^{+}/R_{k,q}^{-} = c^{q} R_{k,0}^{+}/R_{k,0}^{-}$ where $c$ is a constant, for the Ising Glauber this is the case with $c=e^{4 J / T}$. We also observe that the AME and the PA offer results for the susceptibility which are indistinguishable at the resolution of the figure. Although the AME and PA agree at the deterministic and fluctuation level, the finite size corrections to the average values $N^{-1} \langle \pmb{b} \rangle$ are only accurate for the AME, confirming the results obtained for the SIS model.

\begin{figure}[t!]
\includegraphics[width=0.45\textwidth]{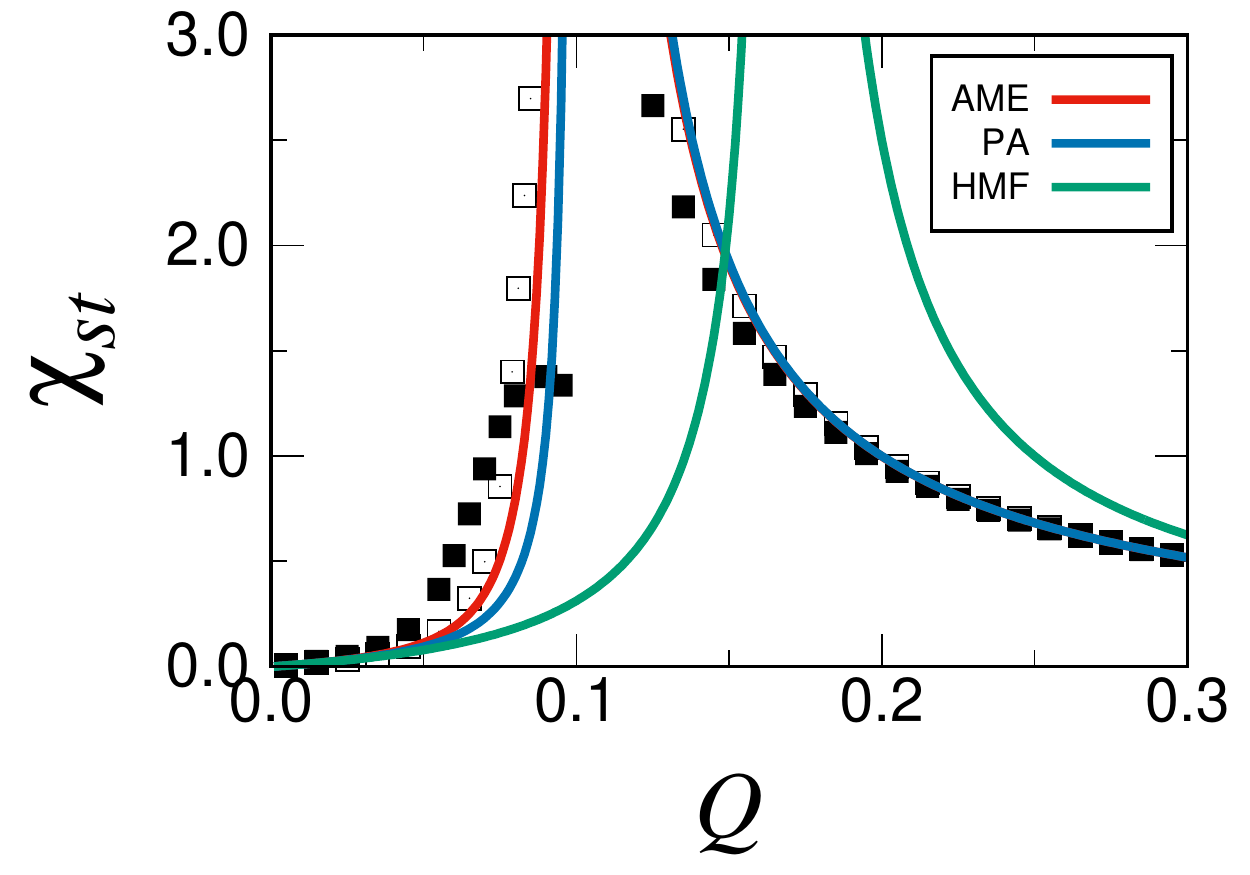}
\includegraphics[width=0.45\textwidth]{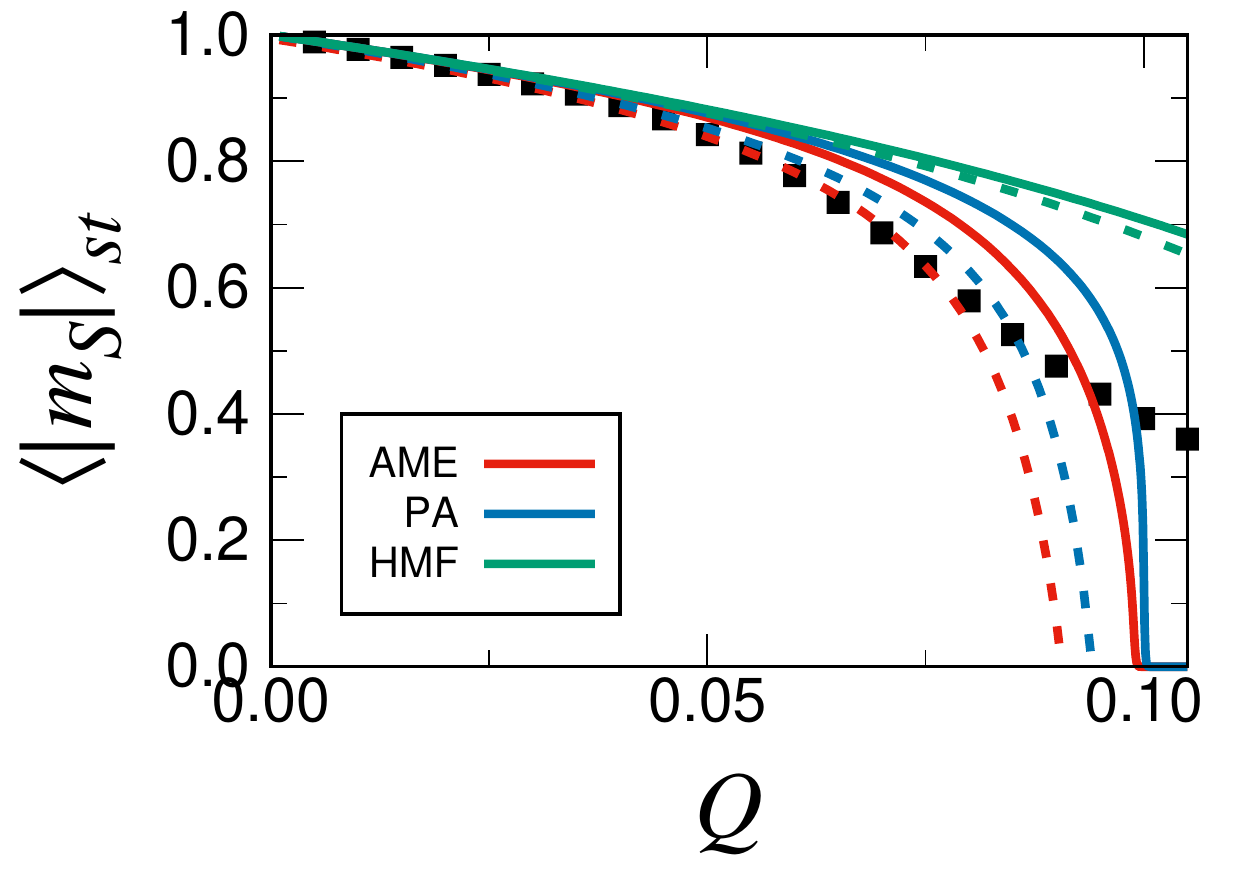}
\caption{Stationary susceptibility $\chi_{\text{st}}$ and average magnetization $\langle \vert m_{S} \vert\rangle_{\text{st}}$ as a function of $Q$ on a 3-regular random network for the majority-vote model. Points correspond to numerical simulations of the model with $N=100$ (solid squares) and $N=400$ (empty squares) averaged over an ensemble of $100$ networks. Lines of different colors are the theoretical prediction of the different approximations (the curves that do not appear are superposed). In the left panel the solid lines are the van Kampen result Eq. (\ref{susceptibility_vK}), while in the right panel the solid line is the deterministic approach and the dashed lines the corrected average values Eq. (\ref{average_values}).}
\label{fig:Majority}
\end{figure}

The third model is the majority-vote model \cite{deOliveira:1992} on a $z$-regular network with rates $R_{k,q}^{+}=Q$ if $q < k/2$, $R_{k,q}^{+}=1/2$ if $q = k/2$, $R_{k,q}^{+}=1-Q$ if $q > k/2$, and $R_{k,q}^{-}=R_{k,k-q}^{+}$, where $Q$ is the rate of spontaneous opinion switching. It is also a symmetric model and it has similar phenomenology to the Glauber model with a critical point $Q_{c}$, see Fig. \ref{fig:Majority}. The most notorious difference is that in this case the AME and PA results are very different even at the deterministic level, and thus for this model only the AME gives more accurate results. The reason for this difference is that the rates do not fulfill the microscopic reversibility condition, see \cite{Gleeson:2013}. Note also in Fig. \ref{fig:Majority} that the AME and PA predict similar critical points $Q_{c}$, but the scaling $\langle \vert m_{S} \vert \rangle_{\text{st}} \propto (Q_{c}-Q)^{\beta}$ is $\beta=1/2$ for the AME and $\beta=1/4$ for the PA. In fact, according to our discussion in the next section, the scaling behavior of the magnetization and susceptibility around a critical point depends on the normal form of the bifurcation. For example, for a typical continuous phase transition as in the Ising model we have $\beta=1/2$ and $\gamma=1$ which corresponds to mean-field exponents. This justifies the common knowledge that critical exponents in complex networks coincide with those of mean-field theory, see \cite{Viana:2004} where the critical exponent of the heat capacity is determined to be $\alpha=0$ (discontinuous heat capacity, which is the mean field result) for the Ising model in a small-world network. Note that this is true as long as the deterministic solution depends on degree moments $\mu_{m}$ that are well defined in the thermodynamic limit $N \rightarrow \infty$. This may not be the case on scale free networks \cite{Gleeson:2013}, where the deterministic solution may depend on the $\mu_{2,3,4,\dots}$ degree moments that diverge, depending on the value of the exponent of the power law degree distribution, as $N \rightarrow \infty$. In this case, this may imply that the critical exponents depend on the details of the degree distribution \cite{Dorogovtsev:2008}. As explained in the next section, one may redefine the finite-size scaling functions and critical exponents to take into account the $N$-dependence of the degree moments.

In the next section we propose a different method for solving the master equation close to a critical region, where the van Kampen expansion fails. We also show how to determine the exponents and scaling properties of the models close to a critical point, for a finite-system and also in the thermodynamic limit $N\rightarrow \infty$.

\section{The expansion around a critical point}\label{sec_critical}

\subsection*{Formulation}
Usually, the rates of the model depend on a set of parameters. Take, for example, a single parameter $T$ for simplicity. It may happen that at a determined value $T=T_{c}$, one of the eigenvalues of the linearized deterministic dynamics becomes equal to zero $D_{1}=0$, this is called critical or bifurcation point. The proposed system size expansion $\x=N\bphi+N^{1/2}\pmb{a}+N^{0}\pmb{b}$ in this case leads to singular, divergent, results for the correlations and average value corrections \cite{Peralta_moments:2018}. The mathematical divergence of the correlations near the critical point is an accurate description only in the strict thermodynamic limit $N \rightarrow \infty$. When $N$ is finite, near the critical point, we have an anomalous scaling with system size, which implies that we have to consider a different ansatz for the system size expansion \cite{Peralta_pair:2018}.

In order to deal with such situations, we start by finding the linear transformation that diagonalizes the Jacobian matrix $\mathbf{B}_{\text{st}} = \mathbf{P} \mathbf{D} \mathbf{P}^{-1}$ being $\mathbf{D}$ the diagonal matrix composed by the eigenvalues and $\mathbf{P}$ the matrix of change of basis whose columns are the corresponding eigenvectors, all evaluated at the critical point $T=T_{c}$. We define the transformed variables in the eigenvector basis $\mathbf{u}=\mathbf{P}^{-1} \bphi$, such that the deterministic dynamics of the new variables is
\begin{equation}
\label{deterministic_u}
\frac{d \mathbf{u}}{d t} = \mathbf{U} \equiv \mathbf{P}^{-1} \pmb{\Phi} (\mathbf{P} \mathbf{u}).
\end{equation}
At the critical point we have $U_{i}(T_{c},\mathbf{u}_{\text{st}})=0$ and $\partial_{u_{j}}U_{i}(T_{c},\mathbf{u}_{\text{st}}) = -D_{i} \delta_{ij}$ with $D_{1}=0$. The Center Manifold Theory \cite{Fernandez:1985,Oppo:1986,Guckenheimer:2002} states that, in this case, there exists a special trajectory or center manifold $u_{i}=h_{i}(T,u_{1})$ for $i \neq 1$ with $u_{i}^{\text{st}} = h_{i}(T,u_{1}^{\text{st}})$ and $\partial_{u_{1}}h_{i}(T_{c},u_{1}^{\text{st}})=0$, that describes locally the dynamics of $\mathbf{u}$ close to the critical point $T_{c}$ and near the fixed point $\mathbf{u}_{\text{st}}$. This implies that the time dependence of the fast variables $u_{i>1}(t)$ is enslaved to the slow variable $u_{1}(t)$. We can write the dependence of $h_{i}(T,u_1)$ as a series expansion
\begin{eqnarray}
\label{hfunction}
 h_{i}(T,u_1) &=& u_{i}^{\text{st}} + \alpha_{i}^{(10)}(T-T_{c}) + \alpha_{i}^{(02)} (u_{1}-u_{1}^{\text{st}})^2  \notag \\&& + \alpha_{i}^{(11)}(T-T_{c})(u_{1}-u_{1}^{\text{st}}) +\dots,
\end{eqnarray}
where the other terms of the expansion are neglected, for example $(T-T_{c})^2$, $(T-T_{c})(u_{1}-u_{1}^{\text{st}})^2$, etc. The coefficients $\alpha_{i}^{(10)}$, $\alpha_{i}^{(11)}$, $\alpha_{i}^{(02)}$ can be determined expanding the dynamical equation $\dot{u}_{i} = \partial_{u_{1}} h_{i} \cdot \dot{u}_1 = U_{i}(T, u_1, h_{2}, h_{3}, \dots)$, the expressions are displayed in Appendix \ref{app:manifold}.

The dynamics of $u_1$ inside the center manifold is $\dot{u}_1 = U_1(T, u_1, h_2, h_{3}, \dots)$ whose series expansion reads 
\begin{eqnarray}
\label{u1_dyn}
\dot{u}_1 &=& \beta^{(0m)} (u_1 - u_1^{\text{st}})^m \notag\\
&& + \beta^{(1n)} (T-T_{c}) (u_1-u_1^{\text{st}})^n+ \dots,
\end{eqnarray}
where $\beta^{(0m)},\,m\ge2$, and $\beta^{(1n)}, \,n\ge 0$, are the lowest non-zero terms in the expansion in powers of $(u_1 - u_1^{\text{st}})$ and higher-order terms are neglected. The expressions of the first coefficients $\beta^{(10)}$, $\beta^{(11)}$, $\beta^{(02)}$, $\beta^{(03)}$ are given in Appendix \ref{app:manifold}.

Equation (\ref{u1_dyn}) is called the normal form of the bifurcation \cite{Guckenheimer:2002} and depending on the value of the coefficients it characterizes three types of critical points/bifurcations. If $\beta^{(10)} \neq 0$ the bifurcation is a {\slshape saddle node}; while if $\beta^{(10)} = 0$, but $\beta^{(11)}\ne0$, the bifurcation is said to be {\slshape transcritical} for $m$ even, or {\slshape pitchfork} for $m$ odd.

From the normal form one can determine the critical exponent $\beta$. Setting the time derivative of Eq. (\ref{u1_dyn}) equal to zero and keeping in mind that $u_{1}^{\text{st}}$ refers to the fixed point at the critical point $u_{1}^{\text{st}}(T_{c})$ we obtain: \\
(i) Saddle node, $0 = \beta^{(0m)} (u_1^{\text{st}} - u_1^{\text{st}})^m+\beta^{(10)}(T-T_{c})$ leading to $u_1^{\text{st}}(T)-u_1^{\text{st}}(T_{c}) \propto \vert T-T_{c} \vert^{1/m}$.\\
(ii) Transcritical and pitchfork bifurcations, $0=\beta^{(0m)} (u_1^{\text{st}} - u_1^{\text{st}})^m+\beta^{(11)} (u_1-u_1^{\text{st}})(T-T_{c})$ leading to $u_1^{\text{st}}(T)-u_1^{\text{st}}(T_{c}) \propto \vert T-T_{c} \vert^{1/(m-1)}$ ($m$ even for the transcritical and odd for the pitchfork).

 Thus $\beta = 1/m$ for the saddle and $\beta = 1/(m-1)$ for the transcritical and pitchfork bifurcations. Note that in Eq.\eqref{u1_dyn} we only keep the two most important terms of the expansion to study the behavior of the stable fixed point close to the transition, and the others can be neglected. This can be checked introducing the first order result $u_1^{\text{st}}(T)-u_1^{\text{st}}(T_{c}) \propto \vert T-T_{c} \vert^{\beta}$ in the expansion Eq. (\ref{u1_dyn}) and evaluating the order of each term.

Once this is understood, we propose the following system-size expansion based on the results of \cite{Nakanishi:1990,Nakanishi:2000,Peralta_pair:2018}. If we approach the critical point as $(T-T_{c}) \sim N^{-r}$, $0<r<1$ and we define the transformed $y_{i}$ variables as $y_{i}=\sum_{j} P_{ij}^{-1} x_{j}$, then $y_{i}$ follows the center manifold with small deviations of order $N^{1/2}$, while the stochastic part of $y_1$ has an anomalous scaling $N^{\upsilon}$, $1/2 <\upsilon<1$, namely:
\begin{eqnarray}
\label{crit_scaling1}
T &=& T_{c}+N^{-r} \xi_{0}, \\
\label{crit_scaling2}
y_1 &=& N u_1^{\text{st}} + N^{\upsilon} \xi_1, \\
\label{crit_scaling3}
y_{i} &=& N h_{i} \left(T, \frac{y_1}{N} \right) + N^{1/2} \xi_{i}.
\end{eqnarray}
Note that $r$ and $\upsilon$ are parameters to be determined and that fluctuations inside the slow center manifold are assumed to scale differently that fluctuations outside it. Using this change of variables $(T, y_{1}, y_{i>1}) \rightarrow (\xi_{0}, \xi_{1}, \xi_{i>1})$, we can expand the master equation (\ref{master_eq}) in powers of $N$, this is done in detail in Appendix \ref{app:manifold}. During the expansion we determine that for a saddle node, $\beta^{(10)} \neq 0$, it is $r=\upsilon=\frac{m}{m+1}$, while for the transcritical or pitchfork bifurcations, $\beta^{(10)} = 0$, it is $\upsilon=\frac{m}{m+1}$, $r=\frac{m-1}{m+1}$. After the expansion of the master equation we obtain a Fokker-Planck equation for the probability $\Pi(\xi_1;t)$ of the slow variable $\xi_{1}$, which for the transcritical and pitchfork bifurcation reads
\begin{widetext}
\begin{equation}
\label{FP_eq_crit}
 \frac{\partial \Pi(\xi_1;t)}{\partial t} = N^{-\frac{m-1}{m+1}} \frac{\partial}{\partial \xi_1} \left[ - \left( \beta^{(11)} \xi_0 \xi_1 + \beta^{(0m)} \xi_1^{m} \right) \Pi + \frac{1}{2} F_{11} \frac{\partial \Pi}{\partial \xi_1} \right],
\end{equation}
\end{widetext}
with a noise intensity $F_{11} = \sum_{i,j} P_{1i}^{-1} P_{1j}^{-1} G_{ij}$. For the saddle node we obtain the same equation but replacing $\beta^{(11)} \xi_0 \xi_1$ by $\beta^{(10)} \xi_0$ . Note that the equation evolves at a slow time scale $\tau = N^{(m-1)/(m+1)}$, this is known in the literature as critical slowing down. In the stationary state for the transcritical and pitchfork bifurcations we have
\begin{equation}
\label{prob_st}
\Pi_{\text{st}}(\xi_1) \propto \exp \left( \frac{\beta^{(11)} \xi_0}{F_{11}} \xi_1^2 + \frac{2 \beta^{(0m)}}{ (m+1)F_{11}} \xi_{1}^{m+1} \right).
\end{equation}
This corresponds to a Gaussian distribution with a saturation term that the van Kampen approach does not take into account. For the saddle node one should replace $\beta^{(11)} \xi_0 \xi_1^2$ by $2 \beta^{(10)} \xi_0 \xi_1$ and the distribution is no longer Gaussian. Note that if $m$ is even we can not integrate the probability Eq. (\ref{prob_st}) in the entire range of $\xi_{1}$ and we have to restrict it to the ``stable'' zone, where fluctuations are not big enough to drive the dynamics to a zone where the deterministic dynamics is unstable and evolves towards infinity. 

Any moment of the $y_{1}$ variable can be computed integrating the distribution Eq. (\ref{prob_st}), for example the variance:
\begin{eqnarray}
\label{fluct_Nfin}
\sigma^{2}[y_{1}] &\equiv& N^{-1} \left( \langle y_{1}^{2} \rangle - \langle y_{1} \rangle^{2} \right) \notag\\
 &=& N^{2\upsilon-1} \cdot \widetilde{\sigma}^2 \big[ N^{r} (T-T_{c})\big],
\end{eqnarray}
where $\widetilde{\sigma}^2[\xi_{0}]$ is the variance of the $\xi_{1}$ variable. Now, the average values and correlations of the $\mathbf{x}$ variables can be related to the transformed variables $\mathbf{y}$ as:
\begin{eqnarray}
\label{av_xi}
\frac{\langle x_i \rangle}{N} &=& \phi_{i}^{\text{st}} + N^{\upsilon-1} P_{i1} \langle \xi_1 \rangle , \\
\label{fluct_xi}
 N^{-1} \left[ \langle x_i x_{j} \rangle - \langle x_i \rangle \langle x_{j} \rangle \right] &=& P_{i1} P_{j1} \sigma^2[y_{1}],
\end{eqnarray}
and from this it is straightforward to determine $\langle m \rangle_{\text{st}}$, $\langle \rho \rangle_{\text{st}}$ and $\chi_{\text{st}}$ with the definitions given in Section \ref{sec_general}.

In the thermodynamic limit $N\rightarrow \infty$, one can show that the van Kampen result is recovered naturally. Take for example a pitchfork bifurcation with $\langle m_{S} \rangle_{\text{st}}=0$. In this case, according to Eq. (\ref{av_xi},\ref{fluct_xi}) the scaling properties of $\langle \vert m_{S} \vert \rangle_{\text{st}}$ and $\chi_{\text{st}}$ with $N$ are
\begin{eqnarray}
\label{scaling_pitch1}
\langle \vert m_{S} \vert \rangle_{\text{st}} &=& N^{\upsilon-1} \widetilde{m} \left[ N^{r} (T-T_{c}) \right],\\
\label{scaling_pitch2}
\chi_{\text{st}} &=& N^{2 \upsilon-1} \widetilde{\chi} \left[ N^{r} (T-T_{c}) \right],
\end{eqnarray}
where $\widetilde{m}(\xi_{0})$ and $\widetilde{\chi}(\xi_{0})$ are the respective scaling functions determined from Eqs. (\ref{prob_st}-\ref{fluct_xi}). In the limit $N \rightarrow \infty$ of Eqs. (\ref{scaling_pitch1}, \ref{scaling_pitch2}), the argument $\xi_{0} = N^{r} (T-T_{c}) \rightarrow \infty$ (for $T \neq T_{c}$) and if we assume the scaling relations $\widetilde{m} \sim \xi_{0}^{\beta}$ and $\widetilde{\chi} \sim \xi_{0}^{-\gamma}$, with appropriate exponents $\beta$ and $\gamma$ such that $\langle \vert m_{S} \vert \rangle_{\text{st}}$ and $\chi_{\text{st}}$ are $N$-independent, we obtain consistently $\beta=\frac{1-v}{r}=\frac{1}{m-1}$ and $\gamma =\frac{2v-1}{r}=1$. Another quantity that is of great interest and that we will use in the next section is the Binder cumulant, defined as the ratio of moments $U_{4} \equiv 1-\frac{\langle m_{S}^4 \rangle}{3 \langle m_{S}^2 \rangle^2}$. It is easy to show that the scaling of this function is given by:
\begin{equation}
\label{scaling_Binder}
U_{4} = \widetilde{u} \left[ N^{r} (T-T_{c}) \right],
\end{equation}
where $\widetilde{u}(\xi_{0}) = 1-\frac{\langle \xi_{1}^4 \rangle_{\text{st}}}{3 \langle \xi_{1}^2 \rangle_{\text{st}}^2}$ is nothing but the Binder cumulant of the $\xi_{1}$ variable, that can be determined using the probability Eq. (\ref{prob_st}) (note that this is independent of the eigenvector coefficients $P_{i1}$, since they cancel out when computing the ratio of moments).

In the next section we apply this method to the models studied in Section \ref{sec_vKresults} and we check if it corrects the problems of the van Kampen expansion in the critical zone. The SIS model can not be studied using these techniques because the model has an absorbing state for $\varepsilon = 0$, and the noise intensity becomes equal to zero in this case, $F_{11}=0$ (evaluated at the absorbing state and at the critical point $\varepsilon = 0$, $\lambda=\lambda_{c}$) and thus, the stationary probability Eq. (\ref{prob_st}) is ill-defined. The finite-size scaling of this type of epidemic models is non-trivial and requires other type of techniques \cite{Satorras:2015}. We will thus consider only the Ising Glauber and majority-vote models.

The scaling functions $\widetilde{m}$ and $\widetilde{\chi}$ generally depend on the degree moments $\mu_{m}$ which may scale with system-size $N$ in a non-trivial way for certain types of highly heterogeneous networks, such as scale-free. It is possible, depending on the model, to reabsorb this $N-$dependence by redefining the scaling functions Eqs. (\ref{scaling_pitch1}, \ref{scaling_pitch2}), see \cite{Peralta:2018}, and this may imply network dependent critical exponents $\beta$, $\gamma$, see \cite{Dorogovtsev:2008}.

\subsection*{Comparison with numerical simulations}\label{sec_critical_results}

\begin{figure}[h!]
\includegraphics[width=0.45\textwidth]{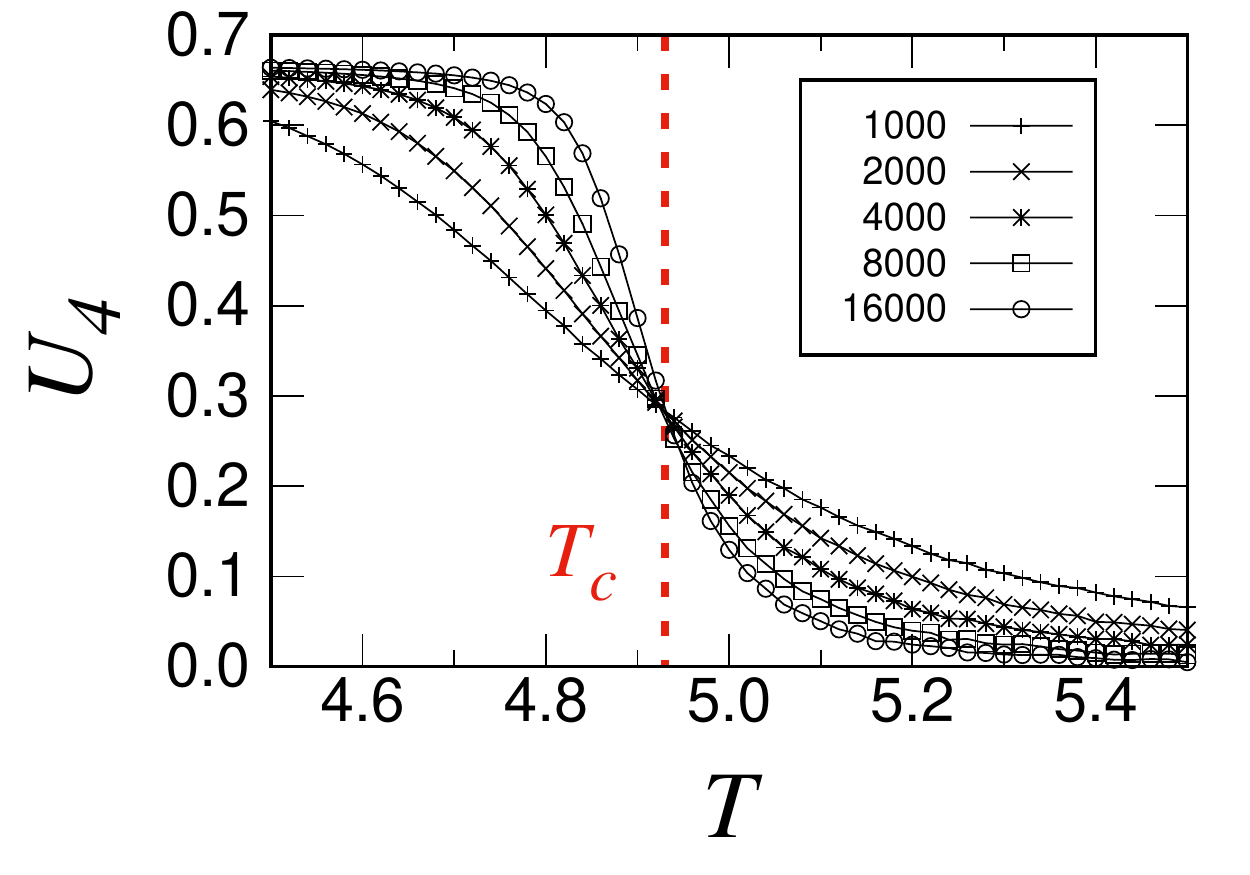}
\includegraphics[width=0.45\textwidth]{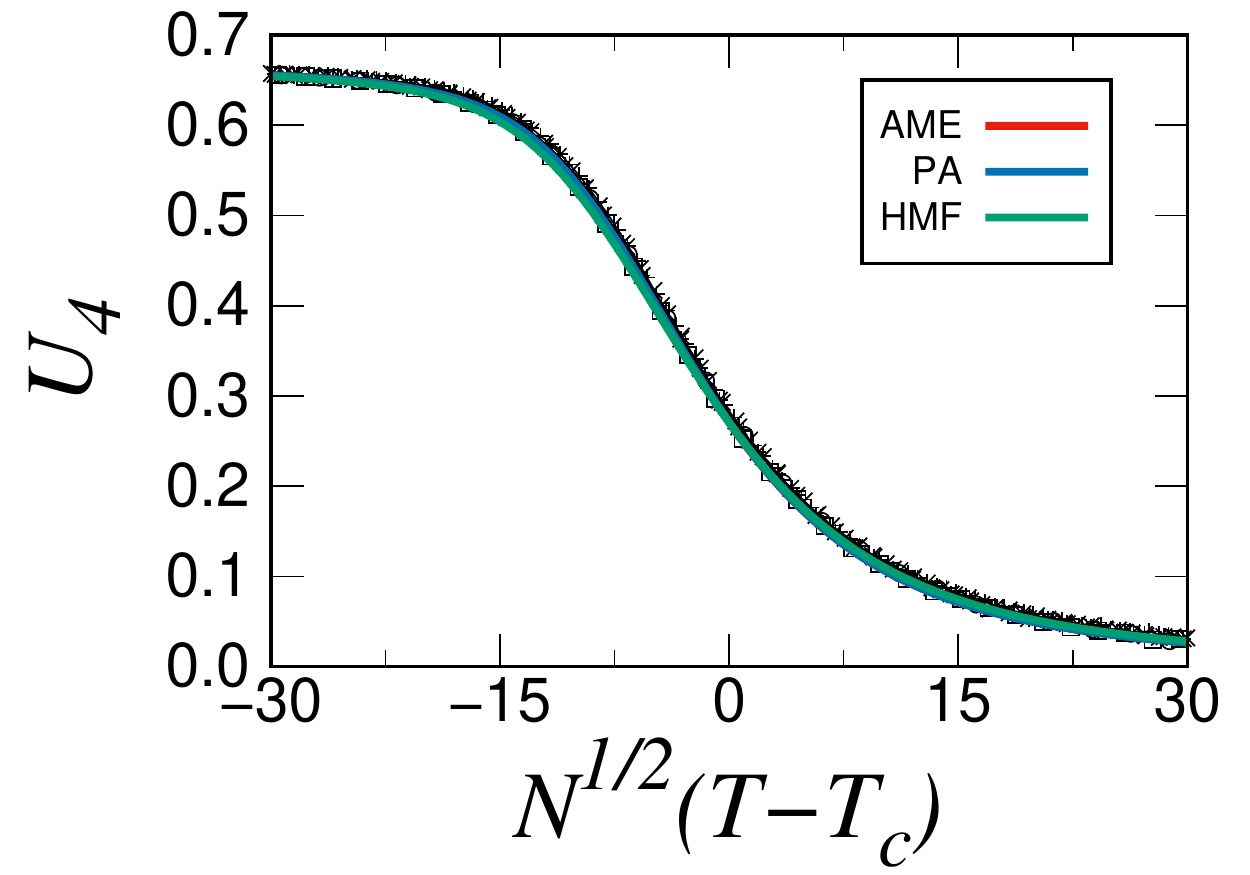}
\caption{Binder cumulant as a function of the temperature $T$ (left panel) and as a function of the rescaled temperature $N^{1/2}(T-T_{c})$ (right panel), for the Ising Glauber model with different system sizes $N$, specified in the figure. The parameters are $J=1$ on an Erd\H{o}s-R\'enyi network with average degree $\mu=5$ and the results were averaged over an ensemble of $100$ networks. Points correspond to numerical simulations of the model with different system sizes $N$ specified in the legend, while lines are the theoretical scaling functions determined from Eq. (\ref{prob_st}, \ref{scaling_Binder}).}
\label{fig:Glauber_binder}
\end{figure}

\begin{figure}[h!]
\includegraphics[width=0.45\textwidth]{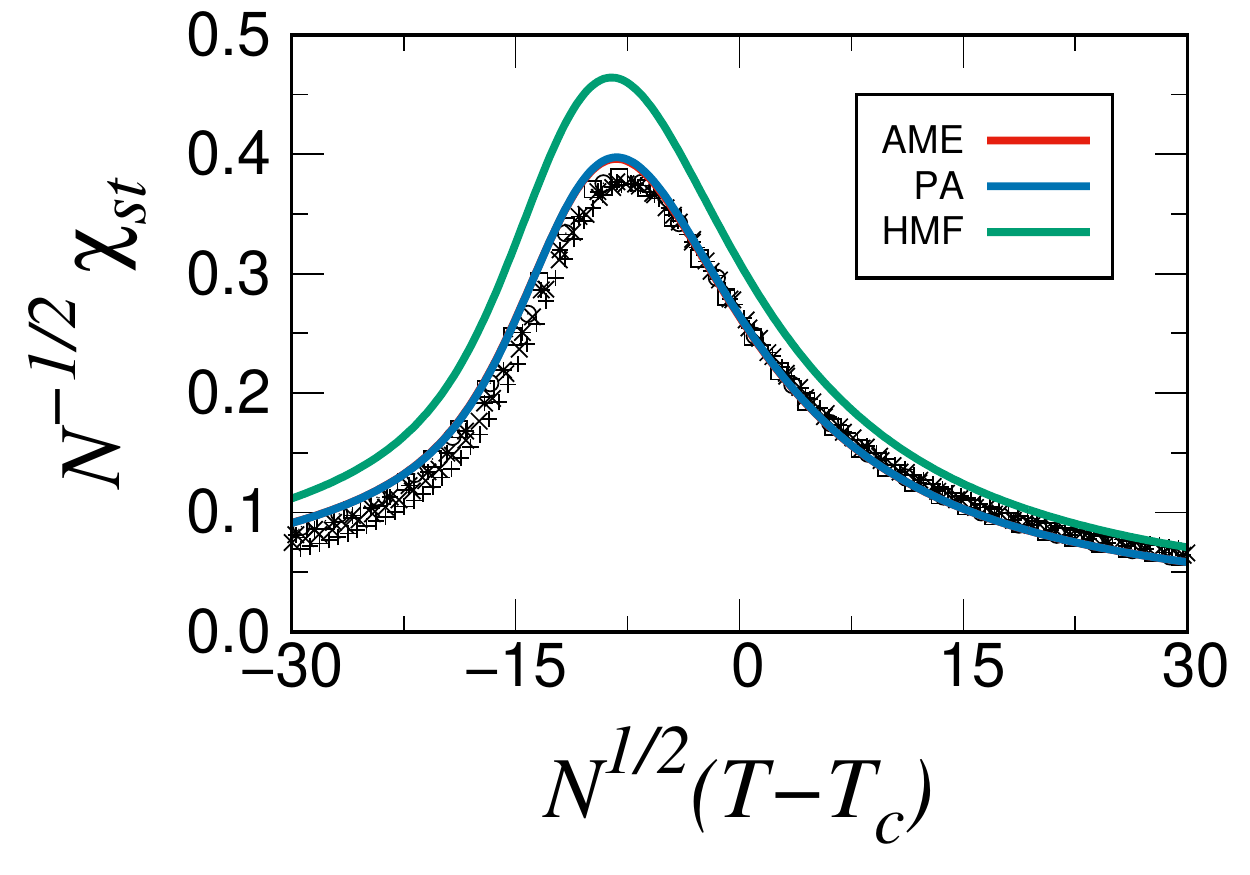}
\includegraphics[width=0.45\textwidth]{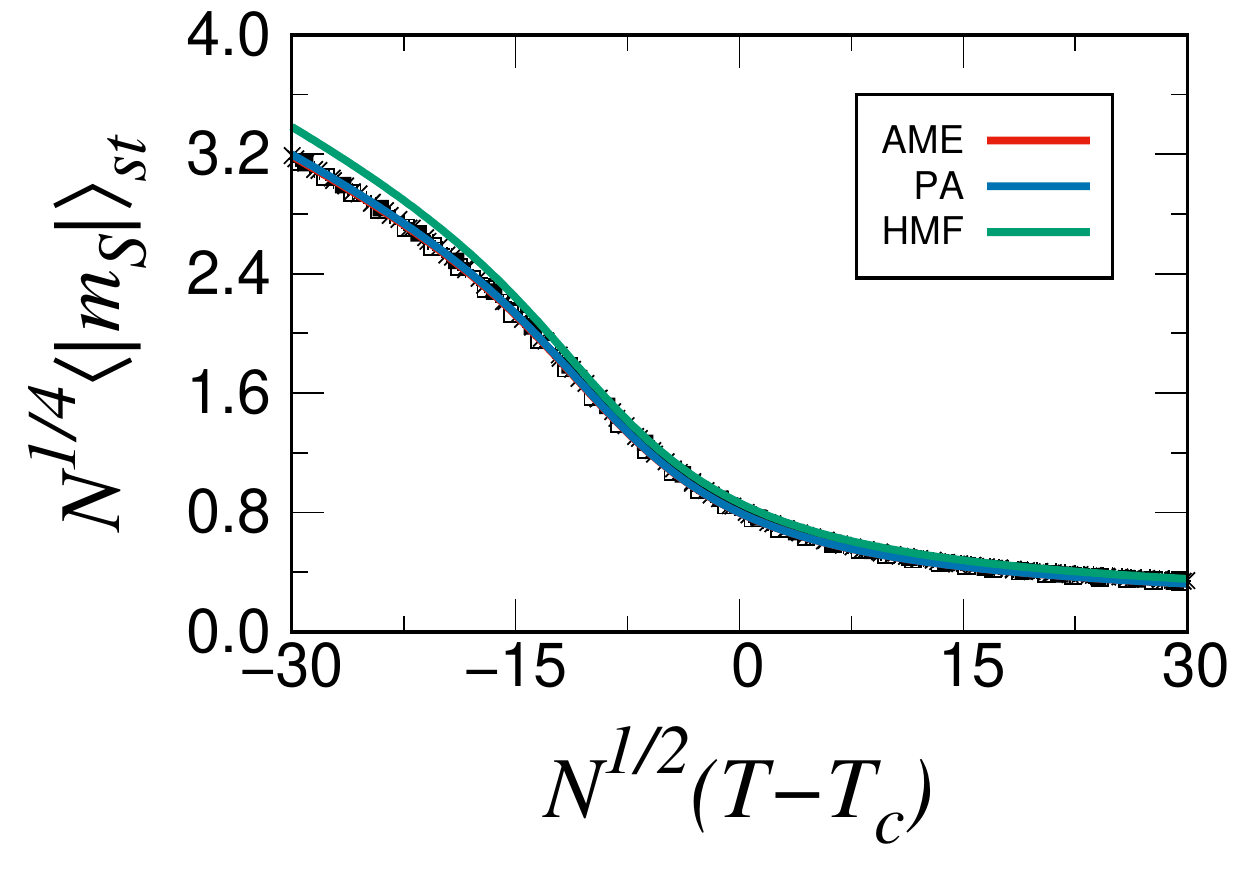}
\caption{Rescaled stationary susceptibility $N^{-1/2} \chi_{\text{st}} = N^{1/2} (\langle m_{S}^2 \rangle_{\text{st}}-\langle \vert m_{S} \vert\rangle^2_{\text{st}})$ and average magnetization $N^{1/4} \langle \vert m_{S} \vert\rangle_{\text{st}}$ as a function of $N^{1/2}(T-T_{c})$ on an Erd\H{o}s-R\'enyi network with average degree $\mu=5$ for the Glauber Ising model with $J=1$. Points correspond to numerical simulations of the model with different system sizes $N$ specified in Fig. \ref{fig:Glauber_binder}, while lines are the theoretical scaling functions determined from Eqs. (\ref{prob_st}, \ref{scaling_pitch1}, \ref{scaling_pitch2}).}
\label{fig:Glauber_scaling}
\end{figure}

We will start with the Ising model with Glauber rates for the network and parameter specifications in the caption of Fig. \ref{fig:Glauber}. The critical point predicted by the AME and PA approximations is $T_{c}(\text{AME}/\text{PA})=4.93\dots$ while for the HMF it is $T_{c}(\text{HMF})=4.96\dots$ In order to determine the critical point numerically from the Monte Carlo (MC) simulations, we use a standard technique of statistical mechanics \cite{Binder:1981}, which consists in computing the Binder cumulant defined as $U_{4} \equiv 1-\frac{\langle m_{S}^4 \rangle}{3 \langle m_{S}^2 \rangle^2}$ for different system sizes $N$, such that all the different curves cross at the critical $T_{c}$, see Fig. \ref{fig:Glauber_binder}. We obtain in this case $T_{c}(\text{MC})=4.93\pm 0.01$ in perfect accordance to the AME/PA results. After computing the coefficients of the normal form of the bifurcation Eq. (\ref{u1_dyn}) we obtain for all three approaches AME/PA/HMF that $\beta^{(10)}=0$, $\beta^{(11)} < 0$, $\beta^{(02)}=0$ and $\beta^{(03)}<0$ which indicates that, according to our discussion in Section \ref{sec_critical}, we have a pitchfork bifurcation with $m=3$, $r=1/2$ and $\upsilon=3/4$. If the theory is correct, and the scaling properties Eqs. (\ref{scaling_pitch1}-\ref{scaling_Binder}) are valid, if we rescale $\langle \vert m_{S} \rangle \rangle_{\text{st}}$, $\chi_{\text{st}}$, $U_{4}$ by $N^{1/4}$, $N^{1/2}$ and $N^{0}$ respectively, and the temperature by $N^{1/2}$, all the curves should collapse on a single universal one $\widetilde{m}(\xi_{0})$, $\widetilde{\chi}(\xi_{0})$ and $\widetilde{u}(\xi_{0})$. In Figs. \ref{fig:Glauber_binder} and \ref{fig:Glauber_scaling} we compute this numerically and compare it with the theoretical scaling functions derived from Eq. (\ref{prob_st}). The matching between numerical and theory is very good which proves the validity of the method. Note also that the scaling functions of the AME and PA coincide, while the HMF shows some deviations. This indicates that there is a strong relation between the validity of the deterministic solution and the scaling functions.

\begin{figure}[h!]
\includegraphics[width=0.45\textwidth]{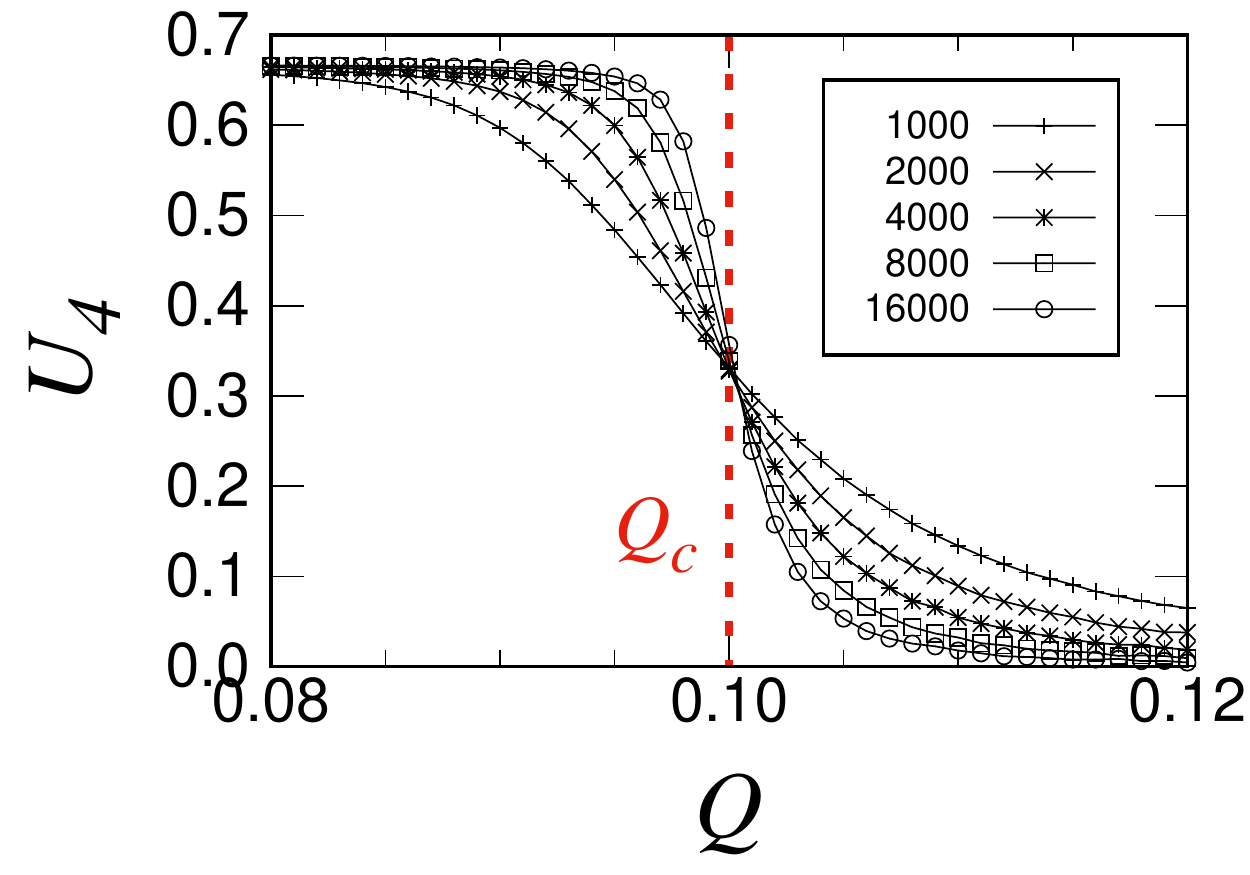}
\includegraphics[width=0.45\textwidth]{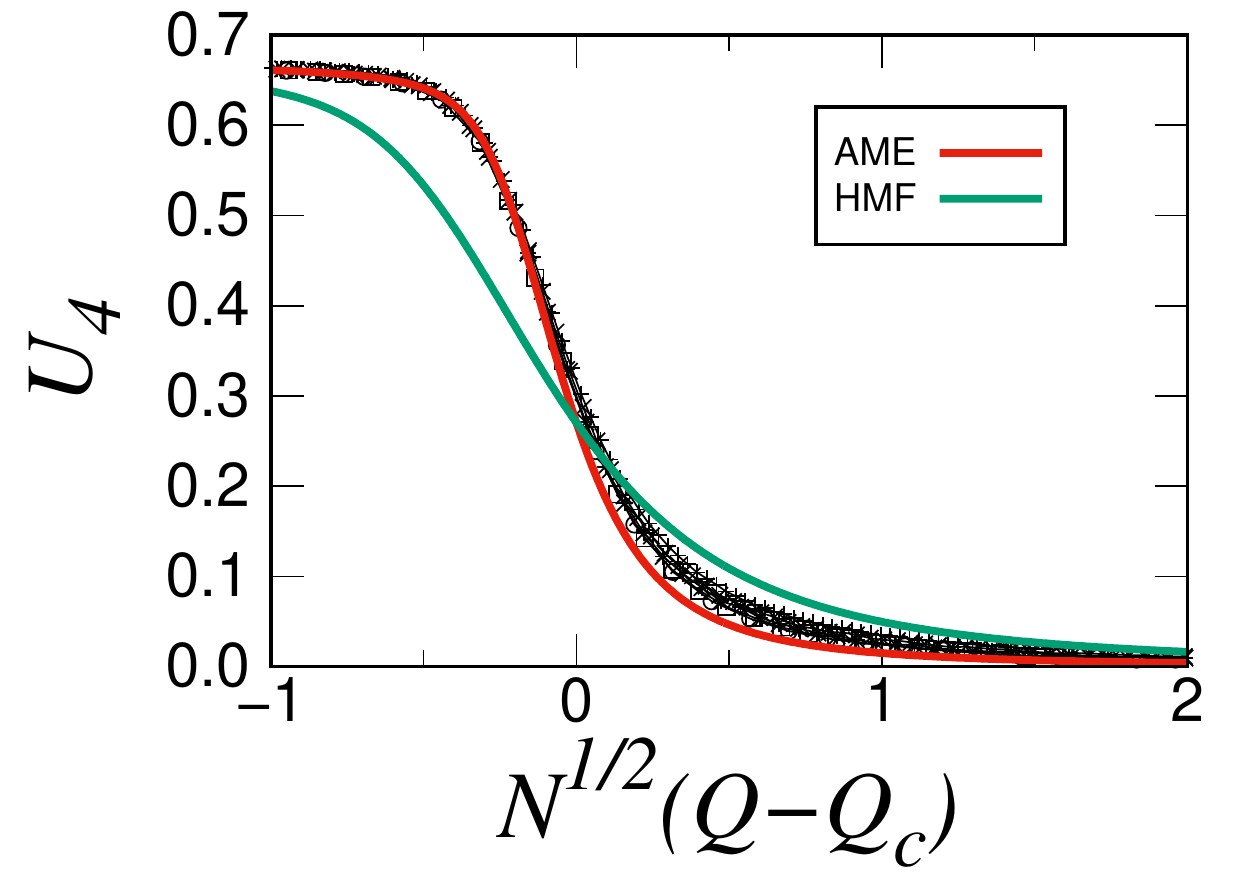}
\caption{Binder cumulant as a function of $Q$ (left panel) and as a function of the rescaled $N^{1/2}(Q-Q_{c})$ (right panel), for the majority-vote model with different system sizes $N$ specified in the legend on a 3-regular random network, and results were averaged over an ensemble of $100$ networks. Points correspond to numerical simulations of the model with different system sizes $N$ specified in the legend, while lines are the theoretical scaling functions determined from Eqs. (\ref{prob_st}, \ref{scaling_Binder}). The finite-size scaling for the PA result is not displayed as it predicts incorrect scaling properties.}
\label{fig:Majority_binder}
\end{figure}

\begin{figure}[h!]
\includegraphics[width=0.45\textwidth]{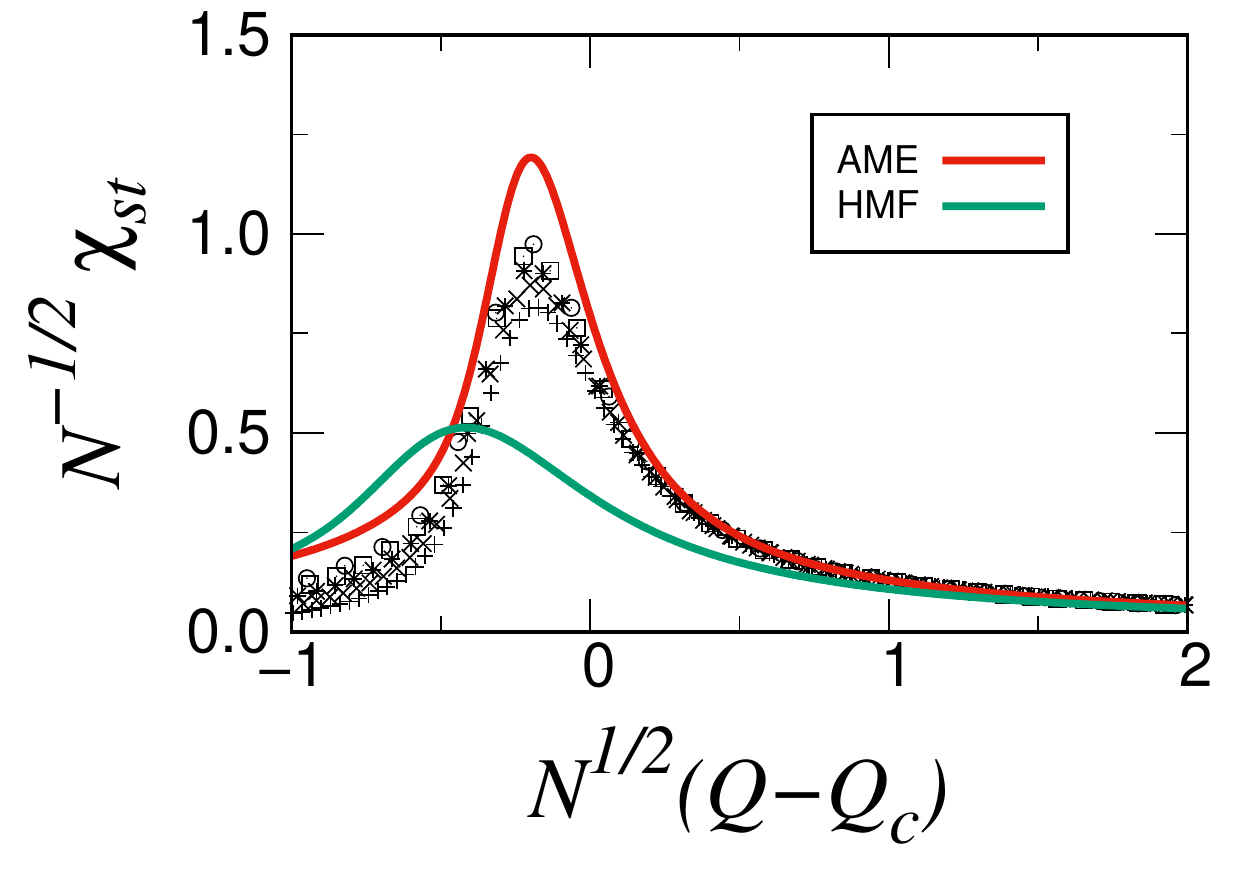}
\includegraphics[width=0.45\textwidth]{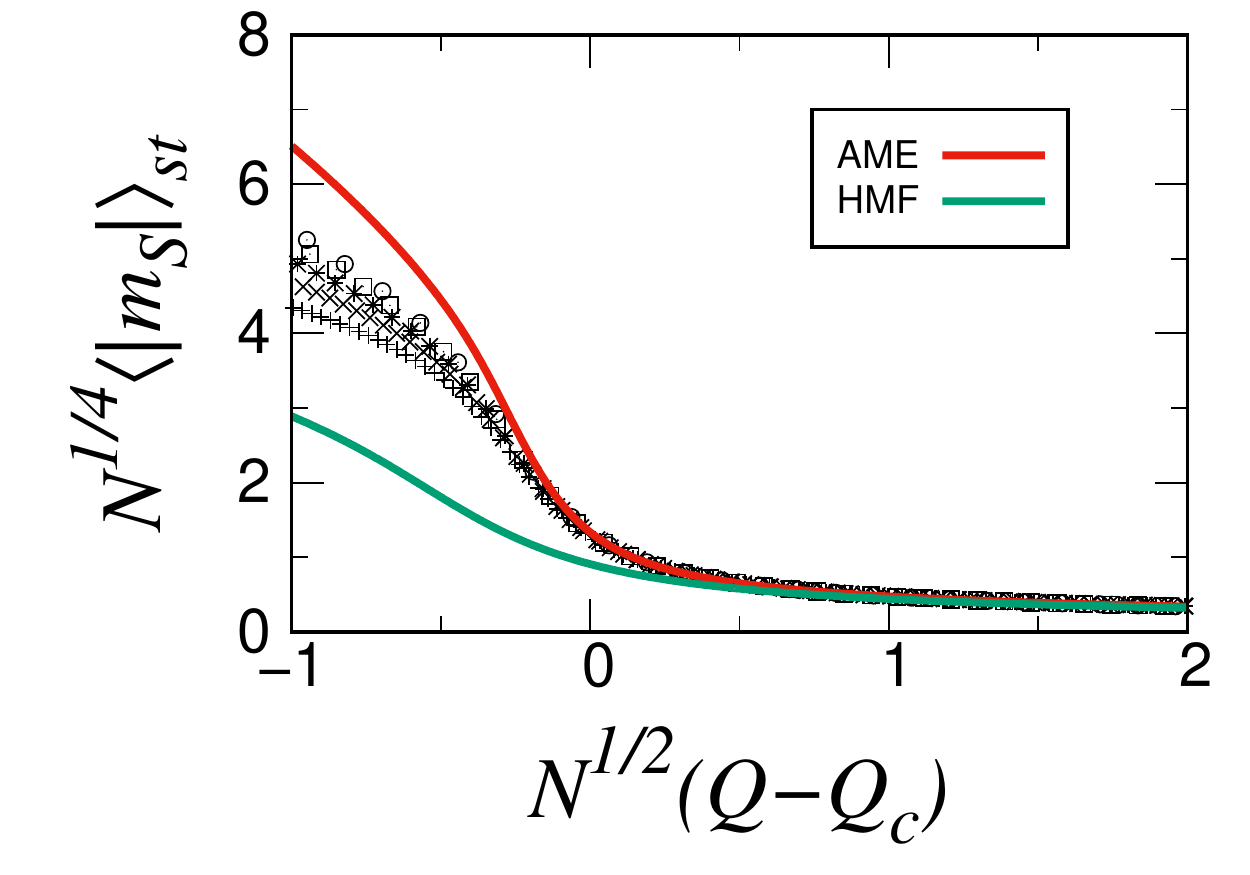}
\caption{Rescaled stationary susceptibility $N^{-1/2} \chi_{\text{st}} = N^{1/2} (\langle m_{S}^2 \rangle_{\text{st}}-\langle \vert m_{S} \vert\rangle^2_{\text{st}})$ and average magnetization $N^{1/4} \langle \vert m_{S} \vert\rangle_{\text{st}}$ as a function of $N^{1/2}(Q-Q_{c})$ on a 3-regular random network for the majority-vote model. Points correspond to numerical simulations of the model with different system sizes $N$ specified in Fig. \ref{fig:Majority_binder}, while lines are the theoretical scaling functions determined from Eq. (\ref{prob_st}, \ref{scaling_pitch1}, \ref{scaling_pitch2}). The finite-size scaling for the PA result is not displayed as it predicts incorrect scaling properties.}
\label{fig:Majority_scaling}
\end{figure}

The next model that we study is the majority-vote model with the same specifications of Fig. \ref{fig:Majority}. The critical point predicted by the AME is $Q_{c}(\text{AME})=0.099\dots$, for the PA it is $Q_{c}(\text{PA})=0.100\dots$ and the HMF is $Q_{c}(\text{HMF})=0.167\dots$. The numerical critical point obtained from the Binder cumulant in Fig. \ref{fig:Majority_binder} is $Q_{c}(\text{MC})=0.100 \pm 0.01$, compatible with the results of the AME and PA but not with the HMF. When we compute the coefficients of the normal form of the bifurcation Eq. (\ref{u1_dyn}) we obtain $\beta^{(10)}=0$, $\beta^{(11)} < 0$, $\beta^{(02)}=0$ and $\beta^{(03)}<0$ for the AME and HMF, which corresponds again to a pitchfork bifurcation with $m=3$, $r=1/2$ and $\upsilon=3/4$. Surprisingly, for the PA we obtain instead $\beta^{(03)}=0$ which suggests a different type of pitchfork with $m=5$, $r=2/3$ and $\upsilon=5/6$. This could be already seen in Fig. \ref{fig:Majority}, as $\langle \vert m_{S} \vert\rangle_{\text{st}} \propto (Q_{c}-Q)^{1/2}$ for the AME but for the PA is more abrupt $\langle \vert m_{S} \vert\rangle_{\text{st}} \propto (Q_{c}-Q)^{1/4}$. As a consequence, we conclude that the PA is not able to capture correctly the scaling properties in this case. In Figs. \ref{fig:Majority_binder} and \ref{fig:Majority_scaling} we compared the theoretical scaling functions with the numerical simulations, where we see that the theoretical scaling of the AME offers a reasonable agreement. Note, however, how the convergence to the theoretical scaling is very slow for $Q<Q_{c}$. This is because for the AME $\beta^{(03)}$ is small, and the other higher order terms of the normal form Eq. (\ref{u1_dyn}) may be important, unless $N$ is extremely large. This also explains the failure of the PA that actually predicts $\beta^{(03)}=0$.

\section{Time dependence}\label{sec_time}

In the previous sections we have focused on stationary averages. The methods, however, are straightforwardly generalized for time dependent results. For the van Kampen approach, we have to solve the deterministic dynamics $\frac{d\pmb{\phi}}{dt}=\pmb{\Phi}$ and, at the same time, the dynamics of the average values and correlations Eqs. (\ref{average_a}-\ref{corr_aa}). On the other hand, if we are close to a critical point in the parameter space, we assume that dynamics evolve following the center manifold and we have to solve Eq. (\ref{FP_eq_crit}), obtaining $\Pi(\xi_{1};t)$. This corresponds to a separation of time scales, which implies that the dynamics outside the manifold is very fast compared to the dynamics inside and thus negligible. We will apply these methods to two different models of interest, not considered in the previous sections, the SI (susceptible-infected) epidemic model and the Threshold model. We chose these models because their dynamics are more interesting than the stationary properties. 

\begin{figure}[h!]
\includegraphics[width=0.45\textwidth]{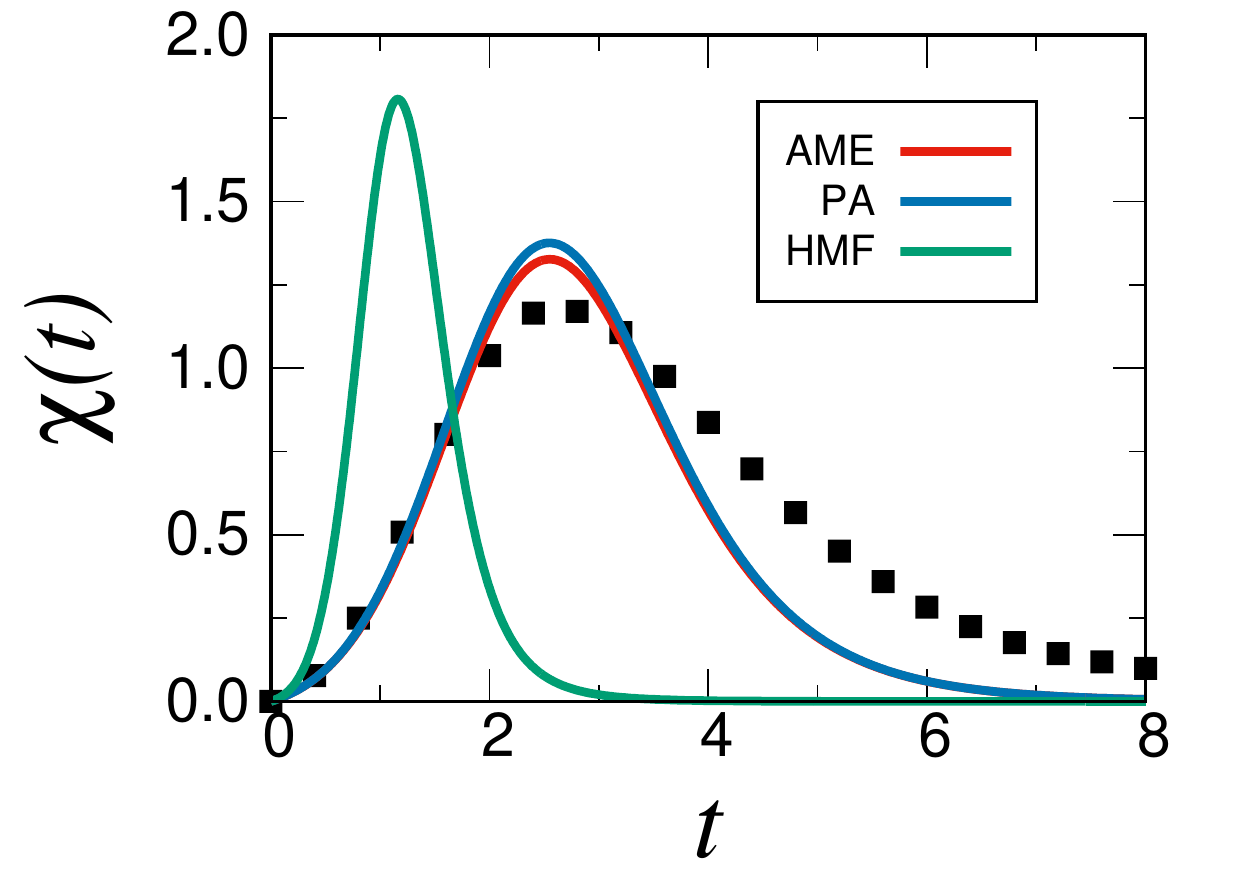}
\includegraphics[width=0.45\textwidth]{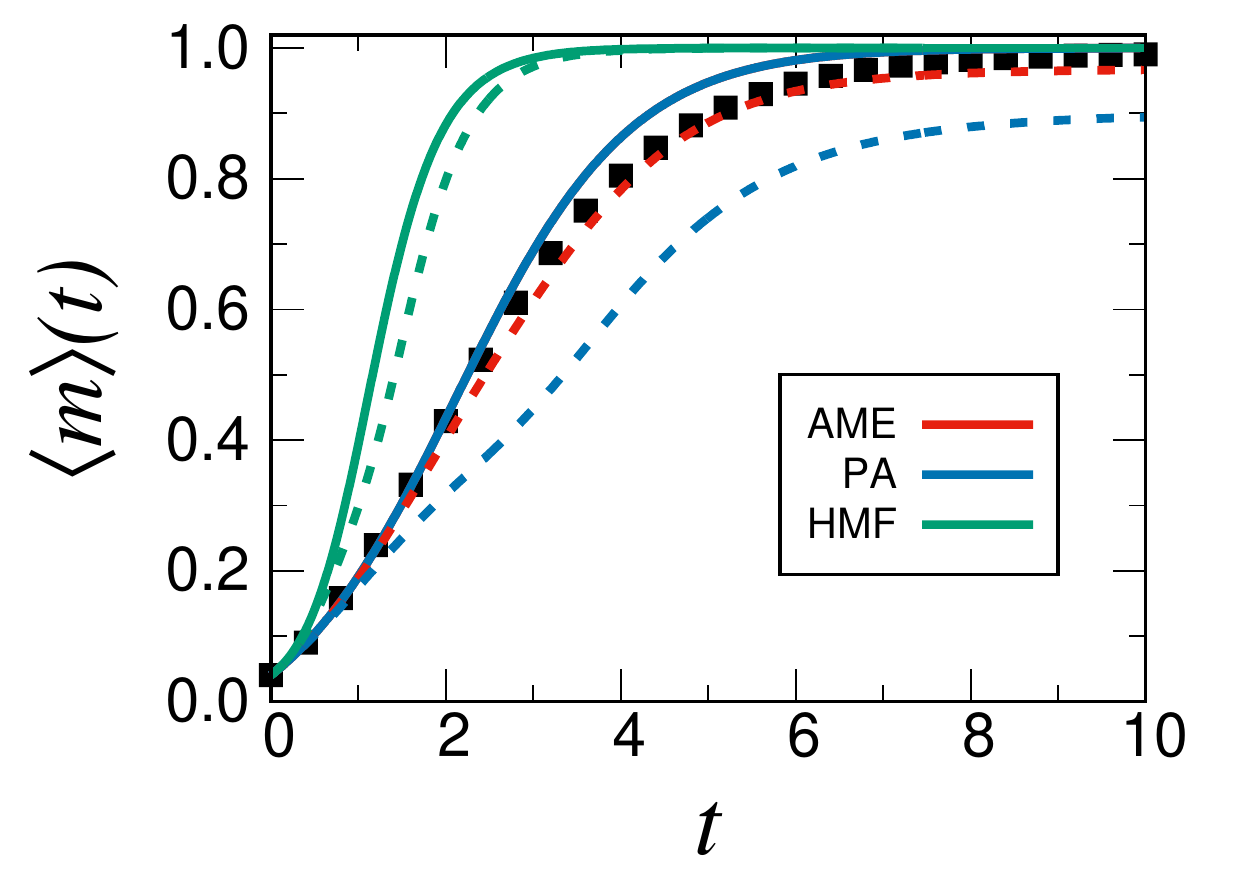}
\caption{Density of active nodes $\langle m(t) \rangle$ and susceptibility $\chi(t)$ as a function of time $t$, for the SI epidemic dynamics with $\lambda=1$ on a scale free network with $P_{k}\sim k^{-2.5}$, $k_{\text{min}}=2$ and $k_{\text{max}}=5$, and $N=25$. Dots are numerical simulations average over $100$ trajectories and $100$ networks, while lines are: (a) in the left panel it is the results of solving the dynamical Eq. (\ref{corr_aa}) for the different approaches, (b) in the right panel the solid lines are the deterministic result $\pmb{\phi}(t)$, while the dashed are corrected by the second order term $\pmb{\phi}(t) + \langle \pmb{b} \rangle(t)/N$ Eq. (\ref{average_b}).}
\label{fig:SI}
\end{figure}

We start with the application of the van Kampen expansion for the SI epidemic model with rates $R_{k,q}^{+} = \lambda q$ and $R_{k,q}^{-}=0$ for a very small system of $N=25$ nodes, see Fig. \ref{fig:SI}. For the susceptibility $\chi(t)$ the AME and PA give a good approximation with slight differences between both approaches, while the HMF shows important discrepancies. For the average value $\langle m(t) \rangle$, the deterministic AME and PA give the same results, while HMF again shows discrepancies. We observe that, similarly to what happens in the stationary state Fig. \ref{fig:SIS}, although the deterministic part of the AME and PA part is equal, the stochastic corrections happen to be only accurate for the AME approach.

\begin{figure}[h!]
\includegraphics[width=0.45\textwidth]{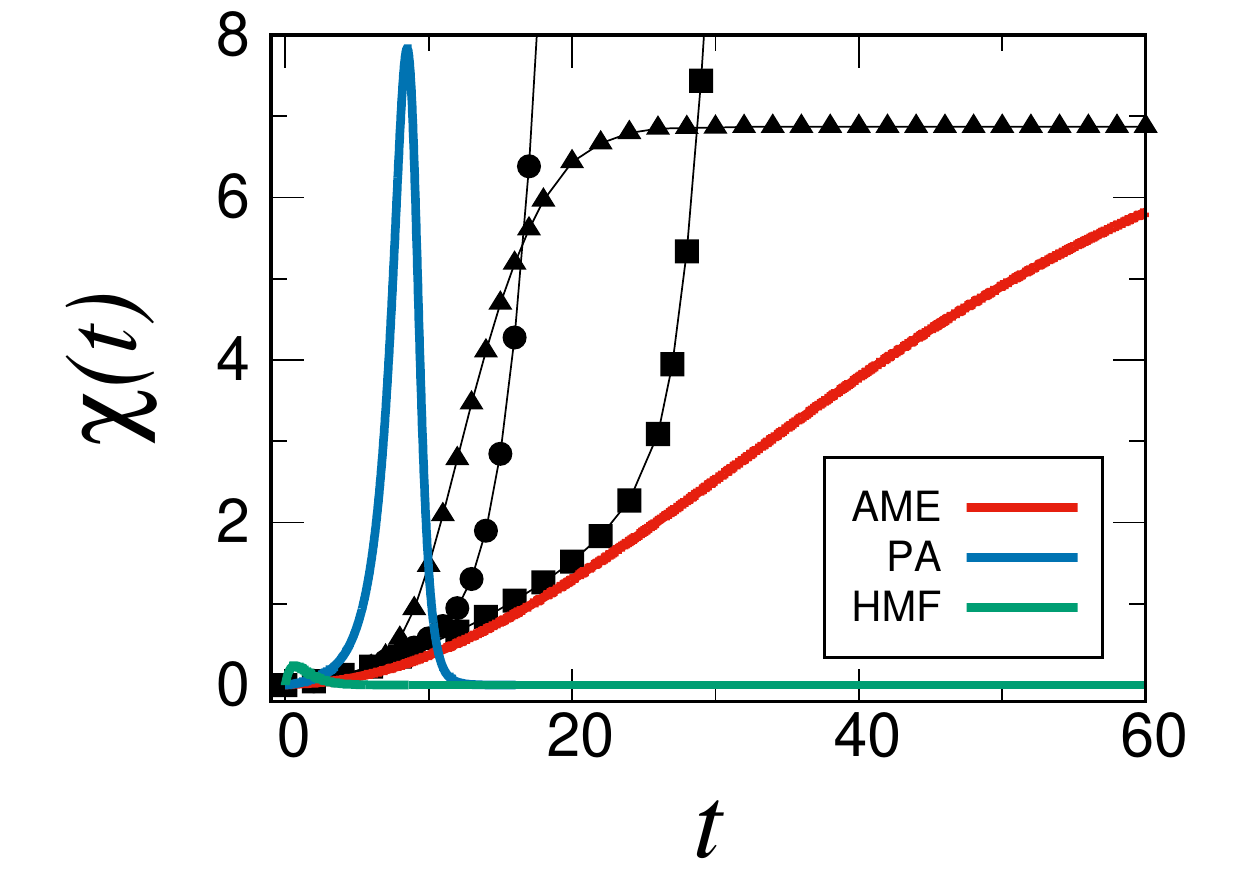}
\includegraphics[width=0.45\textwidth]{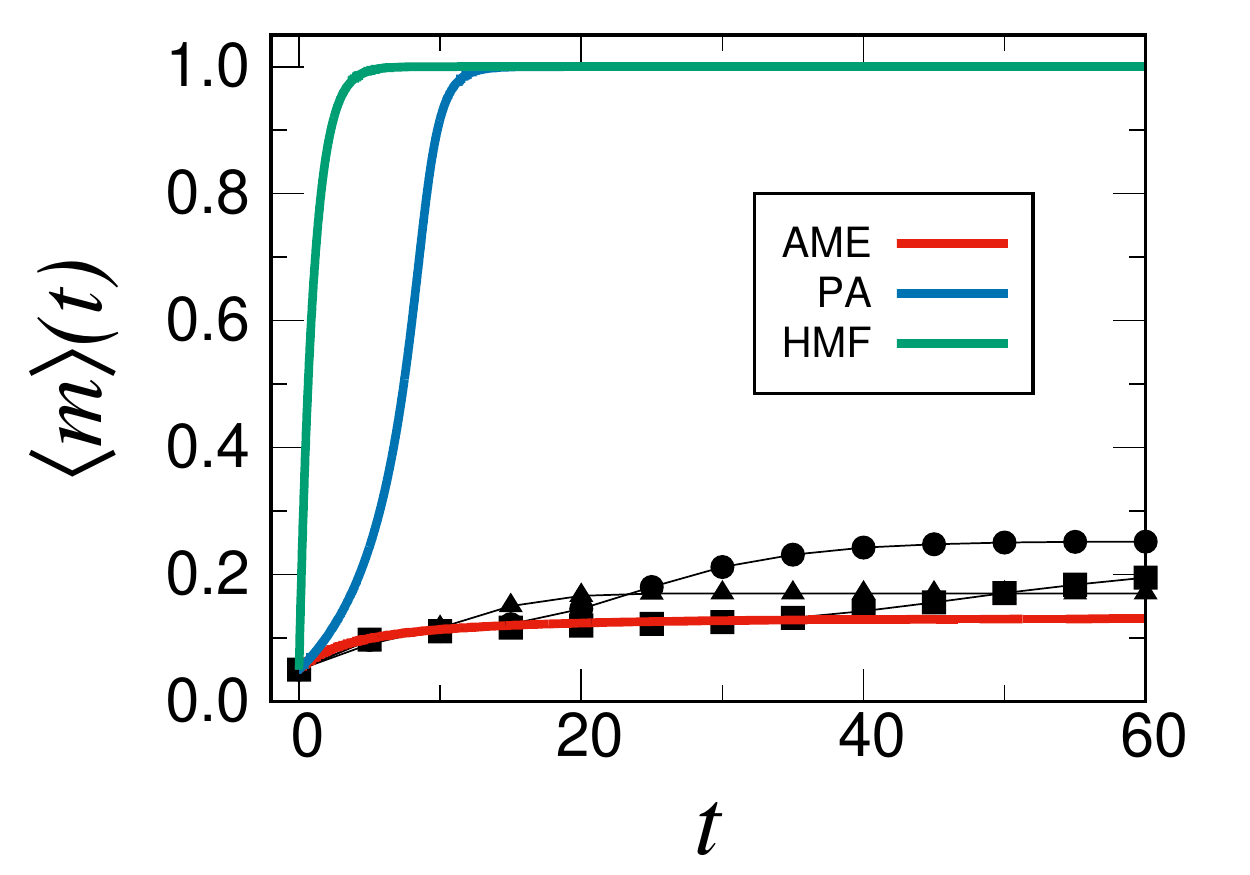}
\caption{Density of active nodes $\langle m(t) \rangle$ and susceptibility $\chi(t)$ as a function of time $t$, for the Threshold model with $M_k=2,\,\forall k$ on a five-regular random network. Dots ($N=100$ triangles, $N=1000$ circles, and $N=10000$ squares) are numerical simulations average over $100$ trajectories and $100$ networks, while lines are: (a) in the left panel it is the results of solving the dynamical Eq. (\ref{corr_aa}) for the different approaches, (b) in the right panel the solid lines are the deterministic result $\pmb{\phi}(t)$.}
\label{fig:Threshold}
\end{figure}

For the Threshold model \cite{Morris:2000,Montanari:2010}, with rates $R_{k,q}^{+}=1$ if $q \geq M_{k}$ and $R_{k,q}^{+}=0$ if $q < M_{k}$ (where $M_{k}$ is a set on integer parameters), $R_{k,q}^{-}=0$, we conclude that the methods presented in this paper are not appropriate for all times $t$. In Fig. \ref{fig:Threshold} we see that the van Kampen system size expansion is accurate only for the AME approach and until a certain time $t<t_{c}$. After that point fluctuations increase with system size and finite size effects become very important. The numerical results of Fig. \ref{fig:Threshold} suggest that $t_{c}$ increases with system size and so do the finite size corrections, at variance with the traditional system size expansion $\langle \pmb{x}(t) \rangle = \pmb{\phi}
(t) + \frac{\langle \pmb{b}(t) \rangle}{N}$. In this special case we should apply alternative methods to solve the Master equation (\ref{master_eq}) for $t > t_{c}$, that are beyond the scope of the current paper.

\section{Summary and Conclusions}\label{sec_conclusion}

In this paper, we have introduced theoretical tools to study stochastic effects in binary-state models on complex networks. First, we constructed the general master equation of the different compartmental approaches: approximate master equation (AME), pair approximation (PA) and heterogeneous mean field (HMF). After that, we elaborated on the different approximate methods for solving the master equation, in particular we explored the van Kampen expansion, valid far from a critical point, and a critical expansion, accurate at the critical zone. From the van Kampen expansion we were able to obtain equations for the correlation matrix of the set of variables and the corrections to their average values, while from the critical expansion we got their finite-size scaling functions. 

We applied these techniques to characterize the stationary properties of the SIS epidemic, Glauber Ising and majority-vote models. When comparing the performance of the different compartmental approaches to numerical simulations we conclude that, if AME and PA have equal or similar results at the deterministic level, the same goes for the fluctuations of the van Kampen expansion and the scaling functions of the critical expansion, but that is not the case for the finite-size corrections to the average values which are only accurate for the AME. This is what we observe for the Glauber model where PA and AME have equivalent deterministic, fluctuations and scaling functions but different finite-size corrections to the average values. This is an indication that, although the PA may work very well in the determination of certain quantities such as average values, the binomial restriction between variables is not necessarily fulfilled by the stochastic trajectories. For the majority-vote model, the AME and PA give different results at all levels, where the PA even predicts incorrectly the scaling coefficients (critical exponents). In general, we can highlight that the probabilistic description using the AME gives very accurate results for stationary and also time-dependent results (as it is shown in Section \ref{sec_time} for the SI epidemic model) within the range of validity of the expansion methods. Certain type of models may not fit in the scope of the expansion methods and their characterization is left for further studies, this is the case of the Threshold model. For $t<t_{c}$ (with $t_c$ increasing with $N$) the van Kampen expansion gives correct results, but for long enough times $t \gg t_{c}$ finite-size effects become more important and deviate from the predicted value.

The solution of the equations for the average values and the fluctuations has been performed numerically using an efficient method developed in \cite{Peralta_moments:2018}. It is left for a future work to explore the possibility of obtaining analytical results of the models and the general conditions for the AME--PA equivalence at the stochastic level. A particularly interesting case that has not been considered in this work is the noisy-voter (Kirman) model for which the linearity of the rates allows one to close the equations for the moments and correlations without the need to resort to the van Kampen approximation. This was done for the homogeneous pair approximation in \cite{Peralta_pair:2018} and it would be interesting to extend these results to the more complicated compartmental models considered in this work. Initial results indicate that the AME corrects the lacks of the PA in specific cases. The differences being specially notorious for low dense -- highly heterogeneous networks \cite{Peralta_inprogress:2020}. 

\section*{Acknowledgements}
Partial financial support has been received from the Agencia Estatal de Investigaci\'on (AEI, MCI, Spain) and Fondo Europeo de Desarrollo Regional (FEDER, UE), under Project PACSS (RTI2018-093732-B-C21/C22) and the Maria de Maeztu Program for units of Excellence in R\&D (MDM-2017-0711). A.F.P. acknowledges support by the Formaci\'on de Profesorado Universitario (FPU14/00554) program of Ministerio de Educaci\'on, Cultura y Deportes (MECD) (Spain). A.F.P. thanks Prof. James Gleeson for the warm hospitality and useful discussions concerning the methods of Ref. \cite{Gleeson:2013} during a stay at the University of Limerick.

\bibliographystyle{unsrt}
\bibliography{Bibliography}
\clearpage

\widetext
\appendix

\section{Expressions of $\mathbf{B}$, $\mathbf{G}$, and Hessian matrices}\label{app:matrices}
In this section, we will calculate all the necessary ingredients involved in the equations of the correlations and corrections Eqs. (\ref{average_a}-\ref{corr_aa}). This essentially includes the Jacobian $\mathbf{B}$-matrix, $\mathbf{G}$-matrix and the set of Hessian matrices $\partial^2_{\phi_{j},\phi_{k}} \Phi_{i}$ that can be later used to determine $\Gamma_{i}= \frac{1}{2} \sum_{j,k} \langle a_{j} a_{k} \rangle \partial^2_{\phi_{j},\phi_{k}} \Phi_{i}$. We will proceed for the different levels of description explained in the main text. All the expressions developed here have been incorporated in a FORTRAN code that can be obtained from the authors upon request.
\subsection*{Approximate Master Equation}
Starting with the variables $\lbrace N_{n,k,q} \rbrace$, the $\mathbf{B}$ matrix can be calculated by taking the derivatives $B_{n,k,q;n',k',q'}=-\partial_{\phi_{n',k',q'}} \Phi_{n,k,q}$, with definitions Eqs. (\ref{Phi0}, \ref{Phi1}), this is:
\begin{eqnarray}
\label{matrixB00}
B_{0,k,q;0,k',q'} &=& \delta_{k,k'} \delta_{q, q'} \left( R_{k,q}^{+} + (k-q) \beta^{s} + q \gamma^{s} \right) - \delta_{k,k'} \delta_{q-1, q'} (k-q+1) \beta^{s} - \delta_{k,k'} \delta_{q+1, q'} (q+1) \gamma^{s} \notag\\
&& + \Big[ (k-q) \phi_{0,k,q}- (k-q+1)\phi_{0,k,q-1} \Big] \frac{\partial \beta^{s}}{\partial \phi_{0,k',q'}}, \\
 \label{matrixB01}
 B_{0,k,q;1,k',q'} &=& -\delta_{k,k'} \delta_{q, q'} R_{k,q}^{-} + \Big[q \phi_{0,k,q} - (q+1) \phi_{0,k,q+1}\Big] \frac{\partial \gamma^{s}}{\partial \phi_{1,k',q'}},\\
 \label{matrixB10}
 B_{1,k,q;0,k',q'} &=& -\delta_{k,k'} \delta_{q, q'} R_{k,q}^{+} + \Big[ (k-q)\phi_{1,k,q} - (k-q+1)\phi_{1,k,q-1} \Big] \frac{\partial \beta^{i}}{\partial \phi_{0,k',q'}},\\
 \label{matrixB11}
 B_{1,k,q;1,k',q'} &=& \delta_{k,k'} \delta_{q, q'} \left( R_{k,q}^{-} + (k-q) \beta^{i} + q \gamma^{i} \right) - \delta_{k,k'} \delta_{q-1, q'} (k-q+1) \beta^{i} - \delta_{k,k'} \delta_{q+1, q'} (q+1) \gamma^{i} \notag\\
&& + \Big[q \phi_{1,k,q} - (q+1)\phi_{1,k,q+1} \Big] \frac{\partial \gamma^{i}}{\partial \phi_{1,k',q'}}.
\end{eqnarray}

Here, the derivatives of the rates Eqs. (\ref{betas}-\ref{gammai}) are:
\begin{eqnarray}
\label{dbetas}
\frac{\partial \beta^{s}}{\partial \phi_{0,k,q}} &=& \frac{(k-q)(R_{k,q}^{+}-\beta^{s})}{\sum_{k,q} (k-q)\phi_{0,k,q}}, \\
\label{dgammas}
\frac{\partial \gamma^{s}}{\partial \phi_{1,k,q}} &=& \frac{(k-q) (R_{k,q}^{-}-\gamma^{s})}{\sum_{k,q} (k-q)\phi_{1,k,q} },\\
\label{dbetai}
\frac{\partial \beta^{i}}{\partial \phi_{0,k,q}} &=& \frac{q (R_{k,q}^{+}-\beta^{i})}{\sum_{k,q} q\phi_{0,k,q} }, \\
\label{dgammai}
\frac{\partial \gamma^{i}}{\partial \phi_{1,k,q}} &=& \frac{ q (R_{k,q}^{-}-\gamma^{i})}{\sum_{k,q} q\phi_{1,k,q} }.
\end{eqnarray}
The $\mathbf{G}$-matrix can be calculated as $G_{n,k,q;n',k',q'}= \sum_{\nu} \ell_{n,k,q}^{(\nu)} \ell_{n',k',q'}^{(\nu)} w^{(\nu)}(\bphi)$, where $\ell_{n,k,q}^{(\nu)}$ are Eqs. (\ref{changes1}-\ref{changes4}) and $w^{(\nu)}(\bphi)$ the intensive version of Eqs. (\ref{rates1}, \ref{rates2}) evaluated at $N_{n,k,q}\rightarrow N \phi_{n,k,q}$. This leads, after lengthy algebra, to:
\begin{align}
\label{matrixG00}
& G_{0,k,q;0,k',q'} = \phi_{0,k,q} R_{k,q}^{+} \delta_{k,k'} \delta_{q,q'} - \phi_{0,k,q} R_{k,q}^{+} (k-q) \left( -P_{0}(0,k',q')+P_{0}(0,k',q'-1) \right) \notag\\
&- \phi_{0,k',q'} R_{k',q'}^{+} (k'-q') \left( -P_{0}(0,k,q)+P_{0}(0,k,q-1) \right) + \beta^{nss} \left[ P_{0}(0,k,q) P_{0}(0,k',q') - P_{0}(0,k,q-1) P_{0}(0,k',q') \right. \notag\\
&- \left. P_{0}(0,k,q) P_{0}(0,k',q'-1) + P_{0}(0,k,q-1) P_{0}(0,k',q'-1) \right] + \beta^{ns} \left[ \delta_{k,k'} \delta_{q,q'} P_{0}(0,k,q) - \delta_{k,k'} \delta_{q-1,q'} P_{0}(0,k,q-1) \right. \notag\\
&- \left. \delta_{k,k'} \delta_{q+1,q'} P_{0}(0,k,q) + \delta_{k,k'} \delta_{q,q'} P_{0}(0,k,q-1) \right] \notag\\
&+ \phi_{1,k,q} R_{k,q}^{-} \delta_{k,k'} \delta_{q,q'} + \phi_{1,k,q} R_{k,q}^{-} (k-q) \left( -P_{1}(0,k',q')+P_{1}(0,k',q'+1) \right) \notag\\
&+ \phi_{1,k',q'} R_{k',q'}^{-} (k'-q') \left( -P_{1}(0,k,q)+P_{1}(0,k,q+1) \right) + \gamma^{nss} \left[ P_{1}(0,k,q) P_{1}(0,k',q') - P_{1}(0,k,q) P_{1}(0,k',q'+1) \right. \notag\\
&- \left. P_{1}(0,k,q+1) P_{1}(0,k',q') + P_{1}(0,k,q+1) P_{1}(0,k',q'+1) \right] + \gamma^{ns} \left[ \delta_{k,k'} \delta_{q,q'} P_{1}(0,k,q) - \delta_{k,k'} \delta_{q-1,q'} P_{1}(0,k,q) \right. \notag\\
&- \left. \delta_{k,k'} \delta_{q+1,q'} P_{1}(0,k,q+1) + \delta_{k,k'} \delta_{q,q'} P_{1}(0,k,q+1) \right],
\end{align}
\begin{align}
\label{matrixG01}
& G_{0,k,q;1,k',q'} = -\phi_{0,k,q} R_{k,q}^{+} \delta_{k,k'} \delta_{q,q'} - \phi_{0,k,q} R_{k,q}^{+} q \left( -P_{0}(1,k',q')+P_{0}(1,k',q'-1) \right) \notag\\
&+ \phi_{0,k',q'} R_{k',q'}^{+} (k'-q') \left( -P_{0}(0,k,q)+P_{0}(0,k,q-1) \right) + \beta^{nsi} \left[ P_{0}(0,k,q) P_{0}(1,k',q') - P_{0}(0,k,q) P_{0}(1,k',q'-1) \right. \notag\\
&- \left. P_{0}(0,k,q-1) P_{0}(1,k',q') + P_{0}(0,k,q-1) P_{0}(1,k',q'-1) \right] -\phi_{1,k,q} R_{k,q}^{-} \delta_{k,k'} \delta_{q,q'} \notag\\
&+ \phi_{1,k,q} R_{k,q}^{-} q \left( -P_{1}(1,k',q')+P_{1}(1,k',q'+1) \right) - \phi_{1,k',q'} R_{k',q'}^{-} (k'-q') \left( -P_{1}(0,k,q)+P_{1}(0,k,q+1) \right) \notag\\
&+ \gamma^{nsi} \left[ P_{1}(0,k,q) P_{1}(1,k',q') - P_{1}(0,k,q) P_{1}(1,k',q'+1) \right. \notag\\
&- \left. P_{1}(0,k,q+1) P_{1}(1,k',q') + P_{1}(0,k,q+1) P_{1}(1,k',q'+1) \right],
\end{align}
\begin{align}
\label{matrixG11}
& G_{1,k,q;1,k',q'} = \phi_{0,k,q} R_{k,q}^{+} \delta_{k,k'} \delta_{q,q'} + \phi_{0,k,q} R_{k,q}^{+} q \left( -P_{0}(1,k',q')+P_{0}(1,k',q'-1) \right) \notag\\
&+ \phi_{0,k',q'} R_{k',q'}^{+} q' \left( -P_{0}(1,k,q)+P_{0}(1,k,q-1) \right) + \beta^{nii} \left[ P_{0}(1,k,q) P_{0}(1,k',q') - P_{0}(1,k,q-1) P_{0}(1,k',q') \right. \notag\\
&- \left. P_{0}(1,k,q) P_{0}(1,k',q'-1) + P_{0}(1,k,q-1) P_{0}(1,k',q'-1) \right] + \beta^{ni} \left[ \delta_{k,k'} \delta_{q,q'} P_{0}(1,k,q) - \delta_{k,k'} \delta_{q-1,q'} P_{0}(1,k,q-1) \right. \notag\\
&- \left. \delta_{k,k'} \delta_{q+1,q'} P_{0}(1,k,q) + \delta_{k,k'} \delta_{q,q'} P_{0}(1,k,q-1) \right] \notag\\
&+ \phi_{1,k,q} R_{k,q}^{-} \delta_{k,k'} \delta_{q,q'} - \phi_{1,k,q} R_{k,q}^{-} q \left( -P_{1}(1,k',q')+P_{1}(1,k',q'+1) \right) - \phi_{1,k',q'} R_{k',q'}^{-} q' \left( -P_{1}(1,k,q)+P_{1}(1,k,q+1) \right) \notag\\
&+ \gamma^{nii} \left[ P_{1}(1,k,q) P_{1}(1,k',q') - P_{1}(1,k,q) P_{1}(1,k',q'+1) - P_{1}(1,k,q+1) P_{1}(1,k',q') + P_{1}(1,k,q+1) P_{1}(1,k',q'+1) \right] \notag\\
&+ \gamma^{ni} \left[ \delta_{k,k'} \delta_{q,q'} P_{1}(1,k,q) - \delta_{k,k'} \delta_{q-1,q'} P_{1}(1,k,q) - \delta_{k,k'} \delta_{q+1,q'} P_{1}(1,k,q+1) + \delta_{k,k'} \delta_{q,q'} P_{1}(1,k,q+1) \right].
\end{align}
The G-matrix is symmetric, so that $G_{1,k,q;0,k',q'}=G_{0,k',q';1,k,q}$. The probabilities Eqs. (\ref{probability00}-\ref{probability11}) must be understood again as evaluated at the deterministic $N_{n,k,q}\rightarrow N \phi_{n,k,q}$, and we have defined the new averaged rates
\begin{eqnarray}
\label{bnss}
\beta^{nss} &=& \sum_{k,q} (k-q)(k-q-1)\phi_{0,k,q} R_{k,q}^{+} , \\
\label{bns}
\beta^{ns} &=& \sum_{k,q} (k-q)\phi_{0,k,q} R_{k,q}^{+} , \\
\label{gnss}
\gamma^{nss} &=& \sum_{k,q} (k-q)(k-q-1)\phi_{1,k,q} R_{k,q}^{-} , \\
\label{gns}
\gamma^{ns} &=& \sum_{k,q} (k-q)\phi_{1,k,q} R_{k,q}^{-} , \\
\label{bnsi}
\beta^{nsi} &=& \sum_{k,q} q (k-q)\phi_{0,k,q} R_{k,q}^{+} , \\
\label{gnsi}
\gamma^{nsi} &=& \sum_{k,q} q (k-q)\phi_{1,k,q} R_{k,q}^{-} , \\
\label{bnii}
\beta^{nii} &=& \sum_{k,q} q(q-1)\phi_{0,k,q} R_{k,q}^{+} , \\
\label{bni}
\beta^{ni} &=& \sum_{k,q} q\phi_{0,k,q} R_{k,q}^{+} , \\
\label{gnii}
\gamma^{nii} &=& \sum_{k,q} q(q-1)\phi_{1,k,q} R_{k,q}^{-}, \\
\label{gni}
\gamma^{ni} &=& \sum_{k,q} q\phi_{1,k,q} R_{k,q}^{-} .
\end{eqnarray}
Similarly as before, they can be interpreted as the total rate at which links, connecting the first and second neighbors of the central node, change from being $0$-$0$, $0$-$1$ or $1$-$1$ when the neighbor node changes state. Here, the symbol $\beta, \gamma$ reflects the state of the neighbor node, while the first super index $s,i$ reflects the state of the central node and the second super index the sate of the second neighbor.

The Hessians $\frac{\partial^2 \Phi_{n,k,q}}{\partial \phi_{n',k',q'} \partial \phi_{n'',k'',q''}}$ can be calculated by taking the derivative of Eqs. (\ref{matrixB00}-\ref{matrixB11}) which leads to
\begin{eqnarray}
\label{dd000}
\frac{\partial^2 \Phi_{0,k,q}}{\partial \phi_{0,k',q'} \partial \phi_{0,k'',q''}} &=& \delta_{k,k'} \big[ - \delta_{q,q'} (k-q) + \delta_{q-1,q'} (k-q+1) \big] \frac{\partial \beta^{s}}{\partial \phi_{0,k'',q''}} \notag\\
&+& \delta_{k,k''} \big[ - \delta_{q,q''} (k-q) + \delta_{q-1,q''} (k-q+1) \big] \frac{\partial \beta^{s}}{\partial \phi_{0,k',q'}} \notag\\
&+& \big[- (k-q)\phi_{0,k,q} + (k-q+1)\phi_{0,k,q-1} \big] \frac{\partial^2 \beta^{s}}{\partial \phi_{0,k',q'} \partial \phi_{0,k'',q''}},\\
\label{dd001}
\frac{\partial^2 \Phi_{0,k,q}}{\partial \phi_{0,k',q'} \partial \phi_{1,k'',q''}} &=& \delta_{k,k'} \big[ - \delta_{q,q'} q + \delta_{q+1,q'} (q+1) \big] \frac{\partial \gamma^{s}}{\partial \phi_{1,k'',q''}},\\
\label{dd011}
\frac{\partial^2 \Phi_{0,k,q}}{\partial \phi_{1,k',q'} \partial \phi_{1,k'',q''}} &=& \big[- q \phi_{0,k,q} + (q+1)\phi_{0,k,q+1} \big] \frac{\partial^2 \gamma^{s}}{\partial \phi_{1,k',q'} \partial \phi_{1,k'',q''}},\\
\label{dd100}
\frac{\partial^2 \Phi_{1,k,q}}{\partial \phi_{0,k',q'} \partial \phi_{0,k'',q''}} &=&\big[- (k-q)\phi_{1,k,q} + (k-q+1)\phi_{1,k,q-1} \big] \frac{\partial^2 \beta^{i}}{\partial \phi_{0,k',q'} \partial \phi_{0,k'',q''}},\\
\label{dd101}
\frac{\partial^2 \Phi_{1,k,q}}{\partial \phi_{0,k',q'} \partial \phi_{1,k'',q''}} &=& \delta_{k,k''} \big[ - \delta_{q,q''} (k-q) + \delta_{q-1,q''} (k-q+1) \big] \frac{\partial \beta^{i}}{\partial \phi_{0,k',q'}},\\
\label{dd111}
\frac{\partial^2 \Phi_{1,k,q}}{\partial \phi_{1,k',q'} \partial \phi_{1,k'',q''}} &=&\delta_{k,k'} \big[ - \delta_{q,q'} q + \delta_{q+1,q'} (q+1) \big] \frac{\partial \gamma^{i}}{\partial \phi_{1,k'',q''}} \notag\\
&+& \delta_{k,k''} \big[ - \delta_{q,q''} q + \delta_{q+1,q''} (q+1) \big] \frac{\partial \gamma^{i}}{\partial \phi_{1,k',q'}} \notag\\
&+& \big[- q \phi_{1,k,q} + (q+1)\phi_{1,k,q+1} \big] \frac{\partial^2 \gamma^{i}}{\partial \phi_{1,k',q'} \partial \phi_{1,k'',q''}}.
\end{eqnarray}
The second derivatives of the rates Eqs. (\ref{betas}-\ref{gammai}) are:
\begin{eqnarray}
\label{dd_betas}
\frac{\partial^2 \beta^{s}}{\partial \phi_{0,k',q'}\partial \phi_{0,k'',q''}} &=& \frac{(k'-q')(k''-q'') \left(2 \beta^{s} - R_{k',q'}^{+} - R_{k'',q''}^{+} \right)}{\left( \sum_{k,q} (k-q) \phi_{0,k,q} \right)^2},\\
\label{dd_gammas}
\frac{\partial^2 \gamma^{s}}{\partial \phi_{1,k',q'}\partial \phi_{1,k'',q''}} &=& \frac{(k'-q')(k''-q'') \left( 2 \gamma^{s} - R_{k',q'}^{-} - R_{k'',q''}^{-} \right)}{\left( \sum_{k,q} (k-q) \phi_{1,k,q} \right)^2},\\
\label{dd_betai}
\frac{\partial^2 \beta^{i}}{\partial \phi_{0,k',q'}\partial \phi_{0,k'',q''}} &=& \frac{q' q'' \left( 2 \beta^{i} - R_{k',q'}^{+} - R_{k'',q''}^{+} \right)}{\left( \sum_{k,q} (k-q) \phi_{0,k,q} \right)^2},\\
\label{dd_gammai}
\frac{\partial^2 \gamma^{i}}{\partial \phi_{1,k',q'}\partial \phi_{1,k'',q''}} &=& \frac{q' q'' \left( 2 \gamma^{i} -R_{k',q'}^{-} - R_{k'',q''}^{-} \right)}{\left( \sum_{k,q} (k-q) \phi_{1,k,q} \right)^2}.
\end{eqnarray}

\subsection*{Pair approximation}
The case of the Pair Approximation can be seen as a reduction and change of variables of the previous more complex case, where $\phi_{0,k,q} = (P_k-\phi_k) \text{Bin}_{k,q} \left[ p_{0,k} \right]$, $\phi_{1,k,q} = \phi_k \text{Bin}_{k,q} \left[ p_{1,k} \right]$ and $p_{0,k}=\phi_{0,k}/(k (P_k-\phi_k))$, $p_{1,k}=\phi_{1,k}/(k \phi_k)$. In this way, the Jacobian matrix of the new variables can be calculated using the chain rule as, e.g. $B_{0,k;0,k'} = \sum_{q,q''} q B_{0,k,q;0,k'',q''} \frac{\partial \phi_{0,k'',q''}}{\partial \phi_{0,k'}}$. This procedure leads to the different matrix elements:

\begin{eqnarray}
\label{matrixBkk}
 B_{k;k'} &=& -\delta_{k,k'} \sum_{q} \left[ \frac{\partial \phi_{0,k,q}}{ \partial \phi_{k}} R_{k,q}^{+} - \frac{\partial \phi_{1,k,q}}{ \partial \phi_{k}} R_{k,q}^{-} \right], \\
\label{matrixBk0k}
 B_{k;0,k'} &=& -\delta_{k,k'} \sum_{q} \frac{\partial \phi_{0,k,q}}{ \partial \phi_{0,k}} R_{k,q}^{+}, \\
\label{matrixBk1k}
B_{k;1,k'} &=& \delta_{k,k'} \sum_{q} \frac{\partial \phi_{1,k,q}}{ \partial \phi_{1,k}} R_{k,q}^{-}, \\
\label{matrixB0kk}
B_{0,k;k'} &=& \delta_{k,k'} \sum_{q} \left[ \frac{\partial \phi_{0,k,q}}{ \partial \phi_{k}} q R_{k,q}^{+} - \frac{\partial \phi_{1,k,q}}{ \partial \phi_{k}} q R_{k,q}^{-} \right] - \frac{\partial \beta^{s}}{\partial \phi_{k'}} (k P_k - k \phi_{k} -\phi_{0,k}) + \frac{\partial \gamma^{s}}{\partial \phi_{k'}} \phi_{0,k} + \delta_{k,k'} k \beta^{s}, \\ 
\label{matrixB0k0k}
 B_{0,k;0,k'} &=& \delta_{k,k'} \sum_{q} \frac{\partial \phi_{0,k,q}}{ \partial \phi_{0,k}} q R_{k,q}^{+} - \frac{\partial \beta^{s}}{\partial \phi_{0,k'}} (k P_k - k\phi_{k} -\phi_{0,k}) + \delta_{k,k'}(\beta^{s} + \gamma^{s}), \\
\label{matrixB0k1k}
 B_{0,k;1,k'} &=& -\delta_{k,k'} \sum_{q} \frac{\partial \phi_{1,k,q}}{ \partial \phi_{1,k}} q R_{k,q}^{-} + \frac{\partial \gamma^{s}}{\partial \phi_{1,k'}} \phi_{0,k}, \\
\label{matrixB1kk}
 B_{1,k;k'} &=& -\delta_{k,k'} \sum_{q} \left[ \frac{\partial \phi_{0,k,q}}{ \partial \phi_{k}} q R_{k,q}^{+} - \frac{\partial \phi_{1,k,q}}{ \partial \phi_{k}} q R_{k,q}^{-} \right] - \frac{\partial \beta^{i}}{\partial \phi_{k'}} (k\phi_{k} -\phi_{1,k}) + \frac{\partial \gamma^{i}}{\partial \phi_{k'}} \phi_{1,k} - \delta_{k,k'} k \beta^{i}, \\ 
\label{matrixB1k0k}
 B_{1,k;0,k'} &=& -\delta_{k,k'} \sum_{q} \frac{\partial \phi_{0,k,q}}{ \partial \phi_{0,k}} q R_{k,q}^{+} - \frac{\partial \beta^{i}}{\partial \phi_{0,k'}} (k\phi_{k} -\phi_{1,k}), \\
\label{matrixB1k1k}
 B_{1,k;1,k'} &=& \delta_{k,k'} \sum_{q} \frac{\partial \phi_{1,k,q}}{ \partial \phi_{1,k}} q R_{k,q}^{-} + \frac{\partial \gamma^{i}}{\partial \phi_{1,k'}} \phi_{1,k} + \delta_{k,k'} (\beta^{i} + \gamma^{i}),
\end{eqnarray}
where
\begin{eqnarray}
\label{derivativesPA_0k}
\frac{\partial \phi_{0,k,q}}{\partial \phi_{k}} &=& - \text{Bin}_{k,q} [p_{0,k}] + \text{Bin}_{k,q}'[p_{0,k}] p_{0,k},\\
\label{derivativesPA_1k}
\frac{\partial \phi_{1,k,q}}{\partial \phi_{k}} &=& \text{Bin}_{k,q} [p_{1,k}] - \text{Bin}_{k,q}'[p_{1,k}] p_{1,k},\\
\label{derivativesPA_00k}
\frac{\partial \phi_{0,k,q}}{\partial \phi_{0,k}} &=& \frac{1}{k} \text{Bin}_{k,q}'[p_{0,k}],\\
\label{derivativesPA_11k}
\frac{\partial \phi_{1,k,q}}{\partial \phi_{1,k}} &=& \frac{1}{k} \text{Bin}_{k,q}'[p_{1,k}],
\end{eqnarray}
and the derivative of the binomial distribution is $\text{Bin}_{k,q}'[x] = \frac{q-x k}{x(1-x)} \text{Bin}_{k,q}[x]$.
The derivatives of the rates $\beta^{s}$, $\beta^{i}$, $\gamma^{s}$, $\gamma^{i}$ can be calculated using the chain rule, e.g. $\partial \beta^{s}/\partial \phi_{k} = \sum_{q} ( \partial \beta^{s}/\partial \phi_{0,k,q} ) ( \partial \phi_{0,k,q}/\partial \phi_{k})$.

Similarly, the G-matrix of the Pair approximation can be obtained taking a partial sum of the previous one, e.g. $G_{k;0,k'} = \sum_{q,q'} q' G_{1,k,q;0,k',q'}$. This procedure reads for the different elements:

\begin{eqnarray}
\label{matrixGkk}
G_{k;k'} &=& \delta_{k,k'} \sum_{q} \left[ \phi_{0,k,q} R_{k,q}^{+} + \phi_{1,k,q} R_{k,q}^{-} \right], \\
\label{matrixGk0k}
G_{k;0,k'} &=& -\delta_{k,k'} \sum_{q} \left[q \phi_{0,k,q} R_{k,q}^{+} + q\phi_{1,k,q} R_{k,q}^{-} \right] + \sum_{q}(k-q) \phi_{0,k,q} R_{k,q}^{+} \sum_{q'} P_{0}(0,k',q') \notag\\
 &+& \sum_{q} (k-q)\phi_{1,k,q} R_{k,q}^{-} \sum_{q'} P_{1}(0,k',q'),\\
\label{matrixGk1k}
G_{k;1,k'} &=& \delta_{k,k'} \sum_{q} \left[ q \phi_{0,k,q} R_{k,q}^{+} + q\phi_{1,k,q} R_{k,q}^{-} \right] + \sum_{q}q \phi_{0,k,q} R_{k,q}^{+} \sum_{q'} P_{0}(1,k',q') \notag\\
 &+& \sum_{q} q \phi_{1,k,q} R_{k,q}^{-} \sum_{q'} P_{1}(1,k',q'),\\
\label{matrixG0k0k}
G_{0,k;0,k'} &=& \delta_{k,k'} \sum_{q} \left[ q^2\phi_{0,k,q} R_{k,q}^{+} +q^2 \phi_{1,k,q} R_{k,q}^{-} \right] \\
 &-& \sum_{q} q (k-q)\phi_{0,k,q} R_{k,q}^{+} \sum_{q'} P_{0}(0,k',q') - \sum_{q'} q' (k'-q')\phi_{0,k',q'} R_{k',q'}^{+} \sum_{q} P_{0}(0,k,q) \notag\\
 &-& \sum_{q} q(k-q)\phi_{1,k,q} R_{k,q}^{-} \sum_{q'} P_{1}(0,k',q')-\sum_{q'} q'(k'-q') \phi_{1,k',q'} R_{k',q'}^{-} \sum_{q} P_{1}(0,k,q)\notag\\
 &+& \beta^{nss} \sum_{q} P_{0}(0,k,q) \sum_{q'} P_{0}(0,k',q') + \gamma^{nss} \sum_{q} P_{1}(0,k,q) \sum_{q'} P_{1}(0,k',q')\notag\\
 &+& \delta_{k,k'} \left[ \beta^{ns} \sum_{q} P_{0}(0,k,q) + \gamma^{ns} \sum_{q} P_{1}(0,k,q) \right], \notag\\
\label{matrixG0k1k}
G_{0,k;1,k'} &=& -\delta_{k,k'} \sum_{q} \left[ q^2\phi_{0,k,q} R_{k,q}^{+} + q^2\phi_{1,k,q} R_{k,q}^{-} \right] \\
&-& \sum_{q} q^2\phi_{0,k,q} R_{k,q}^{+} \sum_{q'} P_{0}(1,k',q') + \sum_{q'} q' (k'-q')\phi_{0,k',q'} R_{k',q'}^{+} \sum_{q} P_{0}(0,k,q) \notag\\
 &-& \sum_{q} q^2 \phi_{1,k,q} R_{k,q}^{-} \sum_{q'} P_{1}(1,k',q') + \sum_{q'} q' (k'-q')\phi_{1,k',q'} R_{k',q'}^{-} \sum_{q} P_{1}(0,k,q)\notag\\
 &+& \beta^{nsi} \sum_{q} P_{0}(0,k,q) \sum_{q'} P_{0}(1,k',q') + \gamma^{nsi} \sum_{q} P_{1}(0,k,q) \sum_{q'} P_{1}(1,k',q'),\notag\\
\label{matrixG1k1k}
G_{1,k;1,k'} &=& \delta_{k,k'} \sum_{q} \left[ q^2\phi_{0,k,q} R_{k,q}^{+} + q^2\phi_{1,k,q} R_{k,q}^{-} \right] \notag\\
&+& \sum_{q} q^2\phi_{0,k,q} R_{k,q}^{+} \sum_{q'} P_{0}(1,k',q') + \sum_{q'} (q')^2\phi_{0,k',q'} R_{k',q'}^{+} \sum_{q} P_{0}(1,k,q) \notag\\
 &+& \sum_{q} q^2\phi_{1,k,q} R_{k,q}^{-} \sum_{q'} P_{1}(1,k',q') + \sum_{q'} (q')^2\phi_{1,k',q'} R_{k',q'}^{-} \sum_{q} P_{1}(1,k,q) \notag\\
 &+& \beta^{nii} \sum_{q} P_{0}(1,k,q) \sum_{q'} P_{0}(1,k',q') + \gamma^{nii} \sum_{q} P_{1}(1,k,q) \sum_{q'} P_{1}(1,k',q')\notag\\
 &+& \delta_{k,k'}\left[ \beta^{ni} \sum_{q} P_{0}(1,k,q) + \gamma^{ni} \sum_{q} P_{1}(1,k,q) \right].
\end{eqnarray}

The Hessians can be obtained taking a second derivative of the Eqs. (\ref{matrixBkk}-\ref{matrixB1k1k}), which leads to
\begin{eqnarray}
\label{dkkk}
\frac{\partial^2 \Phi_{k}}{\partial \phi_{k'} \partial \phi_{k''}} &=& \delta_{k,k'}\delta_{k,k''} \sum_{q} \left[ \frac{\partial^2 \phi_{0,k,q}}{ \partial \phi_{k}^2} R_{k,q}^{+} - \frac{\partial^2 \phi_{1,k,q}}{ \partial \phi_{k}^2} R_{k,q}^{-} \right],\\
\label{dkk0k}
\frac{\partial^2 \Phi_{k}}{\partial \phi_{k'} \partial \phi_{0,k''}} &=& \delta_{k,k'}\delta_{k,k''} \sum_{q} \frac{\partial^2 \phi_{0,k,q}}{\partial \phi_{k} \partial \phi_{0,k}} R_{k,q}^{+},\\
\label{dkk1k}
\frac{\partial^2 \Phi_{k}}{\partial \phi_{k'} \partial \phi_{1,k''}} &=& -\delta_{k,k'}\delta_{k,k''} \sum_{q} \frac{\partial^2 \phi_{1,k,q}}{\partial \phi_{k} \partial \phi_{1,k}} R_{k,q}^{-},\\
\label{dk0k0k}
\frac{\partial^2 \Phi_{k}}{\partial \phi_{0,k'} \partial \phi_{0,k''}} &=& \delta_{k,k'}\delta_{k,k''} \sum_{q} \frac{\partial^2 \phi_{0,k,q}}{\partial \phi_{0,k}^2} R_{k,q}^{+},\\
\label{dk0k1k}
\frac{\partial^2 \Phi_{k}}{\partial \phi_{0,k'} \partial \phi_{1,k''}} &=& 0,\\
\label{dk1k1k}
\frac{\partial^2 \Phi_{k}}{\partial \phi_{1,k'} \partial \phi_{1,k''}} &=& -\delta_{k,k'}\delta_{k,k''} \sum_{q} \frac{\partial^2 \phi_{1,k,q}}{\partial \phi_{1,k}^2} R_{k,q}^{-},\\
\label{d0kkk}
\frac{\partial^2 \Phi_{0,k}}{\partial \phi_{k'} \partial \phi_{k''}} &=& -\delta_{k,k'}\delta_{k,k''} \sum_{q} \left[ \frac{\partial^2 \phi_{0,k,q}}{ \partial \phi_{k}^2} q R_{k,q}^{+} - \frac{\partial^2 \phi_{1,k,q}}{ \partial \phi_{k}^2} q R_{k,q}^{-} \right] \notag\\
 &+& \frac{\partial^2 \beta^{s}}{\partial \phi_{k'}\phi_{k''}} (kP_{k} -k \phi_k - \phi_{0,k}) - \frac{\partial^2 \gamma^{s}}{\partial \phi_{k'}\phi_{k''}} \phi_{0,k} - \delta_{k,k''} k \frac{\partial \beta^{s}}{\partial \phi_{k'}} - \delta_{k,k'} k \frac{\partial \beta^{s}}{\partial \phi_{k''}},\\
 \label{d0kk0k}
 \frac{\partial^2 \Phi_{0,k}}{\partial \phi_{k'} \partial \phi_{0,k''}} &=& -\delta_{k,k'}\delta_{k,k''} \sum_{q} \frac{\partial^2 \phi_{0,k,q}}{\partial \phi_{k} \partial \phi_{0,k}} q R_{k,q}^{+} \notag\\
 &+& \frac{\partial^2 \beta^{s}}{\partial \phi_{k'}\phi_{0,k''}} (k P_{k} k -k \phi_k - \phi_{0,k}) 
 - \delta_{k,k''} \left(\frac{\partial \gamma^{s}}{\partial \phi_{k'}} + \frac{\partial \beta^{s}}{\partial \phi_{k'}} \right) - \delta_{k,k'} k \frac{\partial \beta^{s}}{\partial \phi_{0,k''}},\\
 \label{d0kk1k}
 \frac{\partial^2 \Phi_{0,k}}{\partial \phi_{k'} \partial \phi_{1,k''}} &=& \delta_{k,k'}\delta_{k,k''} \sum_{q} \frac{\partial^2 \phi_{1,k,q}}{\partial \phi_{k} \partial \phi_{1,k}} q R_{k,q}^{-} - \frac{\partial^2 \gamma^{s}}{\partial \phi_{k'}\phi_{1,k''}} \phi_{0,k}, \\
 \label{d0k0k0k}
 \frac{\partial^2 \Phi_{0,k}}{\partial \phi_{0,k'} \partial \phi_{0,k''}} &=& -\delta_{k,k'}\delta_{k,k''} \sum_{q} \frac{\partial^2 \phi_{0,k,q}}{\partial \phi_{0,k}^2} q R_{k,q}^{+} \notag\\
 &+& \frac{\partial^2 \beta^{s}}{\partial \phi_{0,k'}\phi_{0,k''}} (k P_{k} k -k \phi_k - \phi_{0,k}) - \delta_{k,k''} \frac{\partial \beta^{s}}{\partial \phi_{0,k'}} - \delta_{k,k'} \frac{\partial \beta^{s}}{\partial \phi_{0,k''}},\\
 \label{d0k0k1k}
 \frac{\partial^2 \Phi_{0,k}}{\partial \phi_{0,k'} \partial \phi_{1,k''}} &=& -\delta_{k,k'} \frac{\partial \gamma^{s}}{\partial \phi_{1,k''}},\\
 \label{d0k1k1k}
 \frac{\partial^2 \Phi_{0,k}}{\partial \phi_{1,k'} \partial \phi_{1,k''}} &=& \delta_{k,k'} \delta_{k,k''} \sum_{q} \frac{\partial^2 \phi_{1,k,q}}{ \partial \phi_{1,k}^2} q R_{k,q}^{-} - \frac{\partial^2 \gamma^{s}}{\partial \phi_{1,k'}\phi_{1,k''}} \phi_{0,k},\\
 \label{d1kkk}
\frac{\partial^2 \Phi_{1,k}}{\partial \phi_{k'} \partial \phi_{k''}} &=& \delta_{k,k'}\delta_{k,k''} \sum_{q} \left[ \frac{\partial^2 \phi_{0,k,q}}{ \partial \phi_{k}^2} q R_{k,q}^{+} - \frac{\partial^2 \phi_{1,k,q}}{ \partial \phi_{k}^2} q R_{k,q}^{-} \right] \notag\\
 &+& \frac{\partial^2 \beta^{i}}{\partial \phi_{k'}\phi_{k''}} (\phi_k k - \phi_{1,k}) - \frac{\partial^2 \gamma^{i}}{\partial \phi_{k'}\phi_{k''}} \phi_{1,k} + \delta_{k,k''} k \frac{\partial \beta^{i}}{\partial \phi_{k'}} + \delta_{k,k'} k \frac{\partial \beta^{i}}{\partial \phi_{k''}},\\
 \label{d1kk0k}
 \frac{\partial^2 \Phi_{1,k}}{\partial \phi_{k'} \partial \phi_{0,k''}} &=& \delta_{k,k'}\delta_{k,k''} \sum_{q} \frac{\partial^2 \phi_{0,k,q}}{\partial \phi_{k} \partial \phi_{0,k}} q R_{k,q}^{+} + \frac{\partial^2 \beta^{i}}{\partial \phi_{k'}\phi_{0,k''}} (\phi_k k - \phi_{1,k}) 
 + \delta_{k,k'} k \frac{\partial \beta^{i}}{\partial \phi_{0,k''}},\\
 \label{d1kk1k}
 \frac{\partial^2 \Phi_{1,k}}{\partial \phi_{k'} \partial \phi_{1,k''}} &=& -\delta_{k,k'}\delta_{k,k''} \sum_{q} \frac{\partial^2 \phi_{1,k,q}}{\partial \phi_{k} \partial \phi_{1,k}} q R_{k,q}^{-} -\delta_{k,k''} \left( \frac{\partial \beta^{i}}{\partial \phi_{k'}} + \frac{\partial \gamma^{i}}{\partial \phi_{k'}} \right) - \frac{\partial^2 \gamma^{i}}{\partial \phi_{k'}\phi_{1,k''}} \phi_{1,k}, \\
 \label{d1k0k0k}
 \frac{\partial^2 \Phi_{1,k}}{\partial \phi_{0,k'} \partial \phi_{0,k''}} &=& \delta_{k,k'}\delta_{k,k''} \sum_{q} \frac{\partial^2 \phi_{0,k,q}}{\partial \phi_{0,k}^2} q R_{k,q}^{+} + \frac{\partial^2 \beta^{i}}{\partial \phi_{0,k'}\phi_{0,k''}} ( \phi_k k - \phi_{1,k}) ,\\
 \label{d1k0k1k}
 \frac{\partial^2 \Phi_{1,k}}{\partial \phi_{0,k'} \partial \phi_{1,k''}} &=& -\delta_{k,k''} \frac{\partial \beta^{i}}{\partial \phi_{0,k'}},\\
 \label{d1k1k1k}
 \frac{\partial^2 \Phi_{1,k}}{\partial \phi_{1,k'} \partial \phi_{1,k''}} &=& -\delta_{k,k'} \delta_{k,k''} \sum_{q} \frac{\partial^2 \phi_{1,k,q}}{ \partial \phi_{1,k}^2} q R_{k,q}^{-} - \frac{\partial^2 \gamma^{i}}{\partial \phi_{1,k'}\phi_{1,k''}} \phi_{1,k} - \delta_{k,k''} \frac{\partial \gamma^{i}}{\partial \phi_{1,k'}} - \delta_{k,k'} \frac{\partial \gamma^{i}}{\partial \phi_{1,k''}},
\end{eqnarray}
where
\begin{eqnarray}
\label{derivativesPA_0kk}
\frac{\partial^2 \phi_{0,k,q}}{\partial \phi_{k}^2} &=& \text{Bin}_{k,q}''[p_{0,k}] \frac{p_{0,k}^2}{P_{k}-\phi_{k}},\\
\label{derivativesPA_0k0k}
\frac{\partial^2 \phi_{0,k,q}}{\partial \phi_{k} \partial \phi_{0,k}} &=& \text{Bin}_{k,q}''[p_{0,k}] \frac{p_{0,k}}{k (P_{k}-\phi_{k})},\\
\label{derivativesPA_00k0k}
\frac{\partial^2 \phi_{0,k,q}}{\partial \phi_{0,k}^2} &=& \text{Bin}_{k,q}''[p_{0,k}] \frac{1}{k^2 (P_{k}-\phi_{k})}, \\
\label{derivativesPA_1kk}
\frac{\partial^2 \phi_{1,k,q}}{\partial \phi_{k}^2} &=& \text{Bin}_{k,q}''[p_{1,k}] \frac{p_{1,k}^2}{\phi_{k}},\\
\label{derivativesPA_1k1k}
\frac{\partial^2 \phi_{1,k,q}}{\partial \phi_{k} \partial \phi_{1,k}} &=& - \text{Bin}_{k,q}''[p_{1,k}] \frac{p_{1,k}}{k\phi_{k}},\\
\label{derivativesPA_11k1k}
\frac{\partial^2 \phi_{1,k,q}}{\partial \phi_{1,k}^2} &=& \text{Bin}_{k,q}''[p_{1,k}] \frac{1}{k^2 \phi_{k}},
\end{eqnarray}
and the second derivative of the binomial distribution is $\text{Bin}_{k,q}''[x] = \frac{-k x (1-x) + (q-x k)(q-x k -1+2 x)}{x^2(1-x)^2} \text{Bin}_{k,q}[x]$.
The second derivatives of the rates $\beta^{s}$, $\beta^{i}$, $\gamma^{s}$, $\gamma^{i}$ can be calculated using the chain rule again, e.g.
\begin{align}
\label{second_chain}
\frac{\partial^2 \beta^{s}}{\partial \phi_{k} \partial \phi_{k'}} =& \sum_{q,q'} \frac{\partial^2 \beta^{s}}{\partial \phi_{0,k,q} \partial \phi_{0,k',q'}} \frac{\partial \phi_{0,k,q}}{\partial \phi_{k}} \frac{\partial \phi_{0,k',q'}}{\partial \phi_{k'}} + \delta_{k,k'} \sum_{q} \frac{\partial \beta^{s}}{\partial \phi_{0,k,q} } \frac{\partial^2 \phi_{0,k,q}}{\partial \phi_{k}^2}.
\end{align}

\subsection*{Heterogeneous mean field}
This is the simplest of the approximations where we reduce variables following $\phi_{0,k,q} = (P_k-\phi_k) \text{Bin}_{k,q} \left[ p \right]$, $\phi_{1,k,q} = \phi_k \text{Bin}_{k,q} \left[ p \right]$ with $p=\sum_{k}\phi_{k} k/\mu$. The Jacobian in this case can be obtained taking the derivative of Eq. (\ref{Phi_HMG}), this is:
\begin{eqnarray}
\label{matrixBkk_HMF}
B_{k;k'} &=& - \sum_{q} \left[ \frac{\partial \phi_{0,k,q}}{ \partial \phi_{k'}} R_{k,q}^{+} - \frac{\partial \phi_{1,k,q}}{ \partial \phi_{k'}} R_{k,q}^{-} \right],\\
\label{derivativesHMF}
 \frac{\partial \phi_{0,k,q}}{\partial \phi_{k'}} &=& -\delta_{k,k'} \text{Bin}_{k,q}[p] + (P_{k}-\phi_{k})\frac{k'}{\mu} \text{Bin}_{k,q}'[p] ,\\
 \frac{\partial \phi_{1,k,q}}{\partial \phi_{k'}} &=& \delta_{k,k'} \text{Bin}_{k,q}[p] + \phi_{k} \text{Bin}_{k,q}'[p] \frac{k'}{\mu}.
\end{eqnarray}
The G-matrix is simply $G_{k;k'}=\sum_{q,q'} G_{1,k,q;1,k',q'}$ which leads to
\begin{equation}
\label{matrixGkk_HMF}
G_{k;k'} = \delta_{k,k'} \sum_{q} \left[ \phi_{0,k,q} R_{k,q}^{+} + \phi_{1,k,q} R_{k,q}^{-} \right].
\end{equation}
The second derivatives can be computed as
\begin{eqnarray}
\label{dkkk2}
\frac{\partial^2 \Phi_{k}}{\partial \phi_{k'} \partial \phi_{k''}} &=& \sum_{q} \left[ \frac{\partial^2 \phi_{0,k,q}}{ \partial \phi_{k'} \partial \phi_{k''}} R_{k,q}^{+} - \frac{\partial^2 \phi_{1,k,q}}{ \partial \phi_{k'} \partial \phi_{k''}} R_{k,q}^{-} \right],\\
\frac{\partial^2 \phi_{0,k,q}}{ \partial \phi_{k'} \partial \phi_{k''}} &=& - \text{Bin}_{k,q}'[p] \left( \delta_{k,k'} \frac{k''}{\mu}+\delta_{k,k''} \frac{k'}{\mu} \right) + \left( P_{k}-\phi_{k} \right) \text{Bin}_{k,q}''[p] \frac{k'}{\mu} \frac{k''}{\mu},\\
\frac{\partial^2 \phi_{1,k,q}}{ \partial \phi_{k'} \partial \phi_{k''}} &=& \text{Bin}_{k,q}'[p] \left( \delta_{k,k'} \frac{k''}{\mu}+\delta_{k,k''} \frac{k'}{\mu} \right) + \phi_{k} \text{Bin}_{k,q}''[p] \frac{k'}{\mu} \frac{k''}{\mu}.
\end{eqnarray}

\newpage
\section{The center manifold}\label{app:manifold}

The coefficients $\alpha_{i}^{(10)}$, $\alpha_{i}^{(11)}$, $\alpha_{i}^{(02)}$ of the center manifold $u_{i} = h_{i}(T,u_{1})$ Eq. (\ref{hfunction}) are: 

\begin{eqnarray}
\label{alpha10}
D_{i} \alpha_{i}^{(10)} &=& \partial_{T} U_{i}, \\
\label{alpha02}
D_{i} \alpha_{i}^{(02)} &=& \frac{1}{2} \partial_{u_{1}}^2 U_{i}, \\
\label{alpha11}
D_{i} \alpha_{i}^{(11)} &=& -\alpha_{i}^{(02)} \partial_{T} U_{1} + \partial_{T u_{1}}^2 U_{i} + \sum_{j \neq 1} \partial^2_{u_{1} u_{j}} U_{i} \alpha_{j}^{(10)}.
\end{eqnarray}
The coefficients $\beta^{(10)}$, $\beta^{(11)}$, $\beta^{(02)}$, $\beta^{(03)}$ of the normal form of the bifurcation Eq. (\ref{u1_dyn}) are:

\begin{eqnarray}
\label{beta10}
\beta^{(10)} &=& \partial_{T} U_{1}, \\
\label{beta02}
\beta^{(02)} &=& \frac{1}{2} \partial_{u_1}^2 U_{1}, \\
\label{beta11}
\beta^{(11)} &=& \partial_{T u_1}^2 U_{1} + \sum_{j \neq 1} \partial^2_{u_{1} u_{j}} U_{1} \alpha_{j}^{(10)}, \\
\label{beta03}
\beta^{(03)} &=& \frac{1}{3!} \partial_{u_1}^3 U_{1} + \sum_{j \neq 1} \partial^2_{u_{1} u_{j}} U_{1} \alpha_{j}^{(02)}.
\end{eqnarray}

Using the change of variables Eq. (\ref{crit_scaling1}-\ref{crit_scaling3}), we can expand the master equation Eq. (\ref{master_eq}) in powers of $N$ and derive a Fokker-Planck equation for the probability of the new variable $\Pi(\xi_{1};t)$. The step operator $\prod_{i=1}^ME_i^{-\ell_i^{(\nu)}} = e^{-\pmb{\ell}^{(\nu)} \cdot \nabla_{\mathbf{x}}}$ in the right hand side of Eq. (\ref{master_eq}) for the $\mathbf{y} = \mathbf{P}^{-1} \mathbf{x}$ variables with $\nabla_{\mathbf{x}} = \mathbf{P}^{-1} \nabla_{\mathbf{y}}$ transforms as $-\mathbf{P}^{-1} \pmb{\ell}^{(\nu)}\cdot \nabla_{\mathbf{y}}+\frac12 ( \mathbf{P}^{-1} \pmb{\ell}^{(\nu)}\cdot \nabla_{\mathbf{y}})^2 + \dots$ Taking into account that the derivatives change like $\partial_{y_i} = N^{-1/2} \partial_{\xi_i}$ and $\partial_{y_1} = N^{-\upsilon} \partial_{\xi_1} - N^{-1/2} \sum_{j} \left( N^{-r} \alpha_{j}^{(11)} \xi_{0} + N^{\upsilon-1} 2 \alpha_{j}^{(20)} \xi_{1} + \dots \right)\partial_{\xi_j}$, and integrating the full equation $\int \left[ \prod_{i\neq 1}d \xi_{i} \right]$ the only terms that remain are, (i) associated to the first derivative $\partial_{\xi_{1}} \left[ \dots \Pi(\xi_{1};t) \right]$ we have:
\begin{equation}
\label{master_eq_xi1_1}
- N^{-\upsilon}\sum_{\nu}\sum_{j} P_{1j}^{-1} \ell_{j}^{(\nu)} W^{(\nu)} = - N^{1-\upsilon-r} \beta^{(10)} \xi_{0}
 - N^{-r} \beta^{(11)} \xi_{0} \xi_{1} - N^{(m-1)(\upsilon-1)} \beta^{(0m)} \xi_{1}^m + \dots
\end{equation}
and (ii) associated to the second derivative $\frac12 \partial^2_{\xi_{1}} \left[ \dots \Pi(\xi_{1};t) \right]$ we have
\begin{equation}
\label{master_eq_xi1_2}
N^{-2\upsilon} \sum_{\nu}\sum_{i j} P_{1i}^{-1} P_{1j}^{-1} \ell_{i}^{(\nu)} \ell_{j}^{(\nu)} W^{(\nu)} = N^{1-2 \upsilon} F_{11} + \dots,
\end{equation}
with $F_{11} = \sum_{i,j} P_{1i}^{-1} P_{1j}^{-1} G_{ij}$. Both terms must be of the same order of $N$, thus we must choose $r$ and $\upsilon$ properly obtaining: $r=\upsilon=\frac{m}{m+1}$ if $\beta^{(10)} \neq 0$ and $\upsilon=\frac{m}{m+1}$, $r=\frac{m-1}{m+1}$ if $\beta^{(10)} = 0$. Putting both terms together we obtain finally the Fokker-Planck equation (\ref{FP_eq_crit}).

\end{document}